\documentclass[11pt,a4paper]{article} 
\pdfoutput=1
\usepackage{style}
\usepackage{amsmath}
\usepackage{amssymb}
\usepackage{amsthm}
\usepackage{psfrag}
\usepackage{graphicx}
\usepackage{hyperref}
\usepackage{feynmp}
\usepackage{color}
\usepackage{bm}
\usepackage{multirow}
\usepackage{mathtools}
\usepackage{subfig}  
\usepackage{caption}

\newcommand{\vx}[1]{\ensuremath{\boldsymbol{#1}}}

\newcommand{\br}[1]{\ensuremath{\bar{#1}}}

\newcommand{\dirdel}{\ensuremath{\delta^{(3)}}}

\DeclarePairedDelimiter\abs{\lvert}{\rvert}
\DeclarePairedDelimiter\corr{\langle}{\rangle}

\newcommand{\be}{\begin{equation}}
\newcommand{\ee}{\end{equation}}
\newcommand{\beqa}{\begin{eqnarray}}
\newcommand{\eeqa}{\end{eqnarray}}

\newcommand\g{\gamma}
\newcommand\D{\mathcal{D}}
\newcommand\G{\Gamma}

\newcommand\s{\sigma}

\renewcommand\a{\alpha}
\renewcommand\b{\beta}
\newcommand{\ve}{\varepsilon}
\renewcommand\l{\vx{l}}

\newcommand{\T}{\Theta}

\newcommand{\q}{{\bf q}}
\renewcommand{\k}{{\bf k}}
\newcommand{\x}{{\bf x}}

\def\e{\eta}
\def\d{\partial}
\newcommand{\bseq}{\begin{subequations}}
\newcommand{\eseq}{\end{subequations}}

\newcommand{\ch}{\mathop{\rm ch}\nolimits}

\renewcommand{\ln}{\mathop{\rm ln}\nolimits}

\newcommand{\intq}{\int[dq]}

\newcommand\msout{\bgroup\markoverwith{\textcolor{red}{\rule[2ex]{2pt}{0.8pt}}}\ULon}

\newcommand{\ks}[1]{{\bf k}_{\sigma(#1)}} 
\newcommand{\ggv}[1]{\br{\Gamma}_{#1}} 
\newcommand{\narg}{{\bf k}_1,..., {\bf k}_n} 
\newcommand{\gxx}[2]{\Gamma^{(#1)}_{#2}}

\newcommand{\gthreearg}{\k_1,\k_2,\k_3}
\newcommand{\sumsig}{\sum_{\sigma}}

\relax

\title{
Time-sliced perturbation theory with primordial non-Gaussianity
and effects of large bulk flows on
inflationary oscillating features}

\author[a]{Anagha Vasudevan\footnote{\texttt{anagha.vasudevan@nbi.ku.dk}}}
\author[b,c]{Mikhail M. Ivanov\footnote{\texttt{mi1271@nyu.edu}}}
\author[d,e,c]{Sergey Sibiryakov\footnote{\texttt{sergey.sibiryakov@cern.ch}}}
\author[f]{Julien Lesgourgues\footnote{\texttt{julien.lesgourgues@physik.rwth-aachen.de}}}

\affiliation[a]{Niels Bohr International Academy and Discovery Center, Niels Bohr Institute, \\
Blegdamsvej 17, DK-2100 Copenhagen, Denmark}
\affiliation[b]{Center for Cosmology and Particle Physics, Department of Physics,
New York University,\\
New York, NY 10003, USA}
\affiliation[c]{Institute for Nuclear Research of the
Russian Academy of Sciences, \\ 
\normalsize \it  60th October Anniversary Prospect, 7a, 117312
Moscow, Russia}
\affiliation[d]{Institute of Physics, Laboratory of Particle Physics and Cosmology, \\ 
 \'Ecole Polytechnique F\'ed\'erale de Lausanne,\\ 
CH-1015, Lausanne, Switzerland}
\affiliation[e]{Theory Department, CERN, \\1 Esplanade des Particules, CH-1211 Gen\`eve 23, Switzerland}
\affiliation[f]{Institute for Theoretical Particle Physics and Cosmology (TTK) RWTH Aachen University,\\ D-52056 Aachen, Germany}

\abstract{
We extend the formalism of \textit{time-sliced perturbation theory} (TSPT) for cosmological large-scale structure 
to include non-Gaussian initial conditions. 
We show that in such a case the TSPT interaction vertices acquire new contributions whose time-dependence 
factorizes for the Einstein-de Sitter cosmology. 
The new formulation is free from spurious infrared (IR) enhancements and reveals 
a clear IR structure of non-Gaussian vertices.
We use the new technique to study the evolution of oscillating features in primordial statistics and show that they are damped due to 
non-linear effects of large bulk
flows. We derive the damping factors for the 
oscillating primordial power spectrum and bispectrum 
by means 
of a systematic IR resummation of relevant Feynman diagrams.
}

\begin{document}

\begin{flushright}
CERN-TH-2019-091\\
INR-TH-2019-012\\
TTK-19-22
\end{flushright}

\maketitle

\section{Introduction} 
\label{sec:intro}

Large scale structure (LSS) provides us with a powerful tool to study the 
dynamical and statistical properties of our universe.
This structure has been formed during the epoch of matter domination and its dynamics on large scales
is governed by dark matter and baryons acting
as single pressureless perfect fluids. 
The three-dimensional distribution of LSS can potentially supersede  
the cosmic microwave background (CMB) measurements 
in the number of Fourier modes available for observations.
On the other hand,
the relevant cosmological information encoded in LSS 
is concealed by the non-linear effects related to gravitational clustering.
Disentangling the cosmological information from
these effects presents a highly non-trivial challenge that must be addressed in order to efficiently use the data of the ongoing and future surveys.

The standard tool to model dark matter clustering in the non-linear regime 
is N-body simulations, which are progressively becoming more robust and accessible.
On the other hand, reaching the accuracy level required by 
future surveys still remains computationally expensive \cite{Schneider:2015yka}.
Besides,
implementing non-minimal cosmological scenarios or initial conditions
(e.g. primordial non-Gaussianity) 
still requires non-trivial extensions of the standard N-body codes.

An alternative way is the development of analytic methods based on perturbation theory \cite{Bernardeau:2001qr}. 
Although limited to a range of large scales, this approach has a
number of advantages.  
First, analytic methods allow for a shorter computational time. Second, 
they give valuable insights into the physics of non-linear clustering. Third, 
they are flexible and can be easily extended to models beyond the standard $\Lambda$CDM cosmology.

One of the goals of future LSS surveys is probing primordial
non-Gaussianity (PNG),
which can shed light on the dynamics of inflation or its alternatives. 
This task is complicated by a number of factors. 
First, generic non-Gaussian phenomena are encoded in higher-order statistics, 
which are hard to measure and interpret. 
Second, the primordial non-Gaussian signal must be disentangled from 
that generated by non-linear gravitational instability even from
initially Gaussian fields. 
Perturbation theory provides a natural and computationally efficient way to do this~\cite{Scoccimarro:2003wn}.
Perturbative calculations 
including primordial non-Gaussianity were performed for the 
power spectrum 
\cite{Taruya:2008pg}
and bispectrum
\cite{Sefusatti:2009qh} of matter and biased tracers. 
The analysis of galaxy bias was later refined in \cite{Assassi:2015fma}, 
see \cite{Desjacques:2016bnm} for a review.
The Effective Field Theory (EFT) of LSS 
\cite{Baumann:2010tm,Carrasco:2012cv}
has been applied to compute 
the power spectrum and bispectrum with non-Gaussian initial conditions 
\cite{Assassi:2015jqa}.  
Based on these developments, an analysis of the ability to extract the
PNG signal from LSS 
was carried out in~\cite{Welling:2016dng,Baldauf:2016sjb,dePutter:2018jqk,Sprenger:2018tdb,Karagiannis:2018jdt}.

A particularly interesting question is existence of oscillatory features in the 
statistics of primordial density fluctuations. 
Such {\it resonance} features naturally appear in models inspired by
the axion monodromy inflation \cite{Silverstein:2008sg} 
where the inflaton potential is modulated by small sinusoidal
oscillations \cite{Flauger:2009ab,Flauger:2010ja}.  
Alternatively, they can be generated by the interaction 
between the inflaton and massive particles, thereby encoding 
information about the particle spectrum at inflation   
\cite{Arkani-Hamed:2015bza,Lee:2016vti,Arkani-Hamed:2018kmz}. 
The latter type of non-Gaussianity has got the name of {\it
  cosmological collider signatures}. 

Oscillating features in the primordial power spectrum have been
constrained 
from the CMB data by the Planck collaboration 
\cite{Ade:2015lrj,Akrami:2018odb}. The prospects to improve these  
constraints using future LSS and 21\,cm surveys were studied in 
\cite{Chen:2016vvw,Ballardini:2016hpi,Palma:2017wxu,Chen:2016zuu,Xu:2016kwz}, 
see also the
recent white paper \cite{Slosar:2019gvt} and references therein.
The Planck collaboration has also performed searches for 
oscillating primordial non-Gaussianity in CMB with the null results
\cite{Akrami:2019izv}. The imprint of resonant non-Gaussianity on the
scale dependent halo bias was discussed in \cite{Cabass:2018roz},
whereas Ref.~\cite{Xu:2016kwz} analyzed the sensitivity of 21\,cm
intensity mapping to this type of PNG. 
Refs.~\cite{MoradinezhadDizgah:2018ssw,Meerburg:2016zdz} presented
forecasts for the reach of LSS
and 21\,cm surveys in probing cosmological collider signatures.
As pointed out in  
Ref.~\cite{Vlah:2015zda}, one expects primordial oscillations in LSS
statistics to get damped at late times due to the non-linear effects
of large bulk flows, similarly to the damping of    
baryon acoustic oscillations (BAO). 
Indeed, the displacement of modes with Fourier momentum $k$ 
caused by non-linear coupling to long modes with momentum $q\ll k$ washes 
out the features with a characteristic period of oscillations in Fourier space $k_{osc}
\ll q$ smaller that the long-mode momentum. 
However, this effect was not discussed beyond the power spectrum.

In this paper we study the non-linear damping of primordial oscillations in the 
inflationary 3-point function (bispectrum). 
We work out a systematic procedure of infrared (IR) resummation that captures the effect of large bulk flows. 
Our method is inspired by techniques developed in the context of the BAO and is based on Time-Sliced Perturbation Theory (TSPT) \cite{Blas:2015qsi}. 
The latter has been proven to be a powerful tool for IR resummation
of the power spectrum and higher-order statistics
\cite{Blas:2016sfa,Ivanov:2018gjr}. 

TSPT is a novel way of performing cosmological perturbation theory utilizing the  
machinery of quantum field theory and statistical physics. 
The key idea of this method is to study the time-dependent probability distribution function
of density and velocity fields, instead of these fields themselves as done in the standard cosmological perturbations theory (SPT).  
The central object is the generating functional for cosmological correlators 
whose time evolution is governed by a Liouville equation of statistical mechanics.
A perturbative expansion of the generating functional leads to Feynman rules 
similar to that of a 3-dimensional Euclidean quantum field theory, in which time plays a role of an external parameter.
This expansion produces equal-time correlation functions of density and velocity at fixed time (redshift) slices, which explains the name to the method.

TSPT deals directly with the statistical quantities, such as equal time n-point correlation functions, and thus 
provides a natural framework for studying non-Gaussian initial conditions.
We will show that the incorporation of such conditions is straightforward
and results in new contributions to the TSPT vertices.
The new contributions take particularly simple form for the Einstein-de Sitter (EdS) cosmology, where time dependence 
of the TSPT vertices factors out. 
We introduce a new convenient diagrammatic technique for the
primordial non-Gau\-ssi\-an contributions. 
As suggested by observations, 
the latter should be treated as small perturbations on top of the standard correlation function corresponding to Gaussian initial conditions. 
The eventual expressions for TSPT correlation functions at fixed order of perturbation theory coincide with the SPT result,
although the individual diagrams are quite different. 
In particular, we will show that the non-Gaussian TSPT vertices do not contain any 
singularities originating from dynamical non-linear mode-coupling. In this sense they are IR-safe. Of course, they may still contain physical IR singularities 
which may be contained in the primordial statistics themselves.

Finally, we show how the new reformulation allows for a systematic IR resummation
of the large bulk flow effects on the evolution of oscillatory
primordial statistics.  
We show that IR resummation in this case closely resembles that of the
BAO for the Gaussian initial  
statistics and leads to a smearing of oscillating features.  
In particular, the relevant contributions to be 
resummed have a graphical representation of \textit{daisy diagrams}.
We derive explicit expressions for the non-linear damping factors of the power spectrum and bispectrum. In the latter case the damping depends on the full
configuration of the wavevectors, including their relative angles.

The paper is organized as follows. 
In Sec.~\ref{sec:prel} we set up conventions and 
discuss the standard treatment of non-Gaussian initial conditions in cosmological
perturbation theory.
In Sec.~\ref{sec:tspt} we recapitulate the Gaussian TSPT framework.
In Sec.~\ref{sec:tsptPNG} we extend it by including the non-Gaussian initial conditions.
Section~\ref{sec:ir} contains the proof of the IR safety of the non-Gaussian TSPT vertices. 
In Sec.~\ref{sec:IRres} we present a framework for IR resummation of non-Gaussian vertices
and apply it to  the non-linear evolution of
the resonant power spectrum and bispectrum from axion monodromy inflation.
We conclude in Sec.~\ref{sec:conclusions}.
Some additional material is given in Appendices.

In the numerical calculations throughout the paper we use the Planck
2018 cosmology \cite{Aghanim:2018eyx} with massless neutrinos. The
linear power spectrum is computed with the Boltzmann code \texttt{CLASS} 
\cite{Blas:2011rf}.

\section{Preliminaries}\label{sec:prel}

In this section we set up the conventions and discuss the standard treatment of 
non-Gaussian initial conditions in LSS. Throughout the paper we will
mostly focus on the non-trivial initial 3-point function
(bispectrum). However, our results can be straightforwardly
generalized to the case of higher non-Gaussian statistics. We will 
comment on this point at the relevant places in the text.

\subsection{Primordial and linear statistics}

We will distinguish between {\em primordial} and {\em linear}
bispectra. 
We adopt the inflationary paradigm and call {\em primordial} the
quantities referring to the curvature perturbation $\zeta$ generated
during inflation. 
Its statistical properties are encoded in the power spectrum and the bispectrum,
\bseq
\begin{align}
& \corr{\zeta(\k)\zeta(\k')}=
(2\pi)^3
\delta^{(3)}_D(\k+\k') P_{\zeta}(k)\,,\\
& \corr{\zeta(\k_1)\zeta(\k_2)\zeta(\k_3)} = 
(2\pi)^3
\delta^{(3)}_D(\k_1+\k_2+\k_3)B_\zeta(\k_1,\k_2,\k_3)\,,
\end{align} 
\eseq
where $\delta^{(3)}_D$ is the 3-dimensional Dirac delta-function.
The power spectrum of the primordial curvature perturbations
can be written as,
\be 
P_{\zeta}(k)\equiv 
\frac{\mathcal{A}^2_\zeta}{k^3}\left(\frac{k}{k_*}\right)^{n_s-1}\,,
\ee
where $k_*$ is a pivot scale, $\mathcal{A}^2_\zeta$ 
is the amplitude and  $(n_s-1)$ is the tilt of the spectrum.
Their mean values from the latest Planck CMB temperature and polarization 
measurements \cite{Aghanim:2018eyx} for $k_*=0.05\,$Mpc$^{-1}$
are\footnote{Note that $\mathcal{A}_\zeta^2$ is related to the scalar
  amplitude $A_s$ used in the Planck analysis by 
 $\mathcal{A}^2_\zeta=2\pi^2 A_s$.}
\be 
 \mathcal{A}^2_\zeta=4.14\cdot 10^{-8}\;,~~~~~~~
 n_s =0.965\,. 
\ee
Let us outline some properties of $B_{\zeta}(\k_1,\k_2,\k_3)$ in
typical inflationary models, see Ref.~\cite{Babich:2004gb} for more details. 
Due to momentum conservation and rotational invariance 
the bispectrum is a function of three independent variables only. 
These can be chosen to be either the three momenta norms $k_1,k_2,k_3$,
or the norms of a pair of momenta, say $k_1,k_2$, and the cosine $\mu_{12}$ 
of the angle between them.
Approximate
scale invariance implies that $B_\zeta$ is 
a homogeneous function of degree $-6$, symmetric in its arguments.
The shapes commonly used in data analysis
are\footnote{We use the Planck conventions \cite{Akrami:2019izv} for the
  bispectrum amplitude $f_{NL}$. Note that the Planck collaboration
  uses the variable $\Phi\equiv 3\zeta/5$, instead of $\zeta$, in the
  analysis of non-Gaussianity. 
} 
local \cite{Bartolo:2001cw},
equilateral \cite{Creminelli:2003iq}, and orthogonal \cite{Senatore:2009gt}, 
\bseq
\begin{align}
& B^{\text{loc}.}_\zeta(k_1,k_2,k_3)=\frac{6}{5}f_{NL}^{\text{loc}.}
\Big(P_\zeta(k_1)P_\zeta(k_2)+P_\zeta(k_2)P_\zeta(k_3)
+P_\zeta(k_3)P_\zeta(k_1)\Big)\,,\\ 
& B^{\text{eq}.}_\zeta(k_1,k_2,k_3)=\frac{18}{5}f_{NL}^{\text{eq}.}
\mathcal{A}_{\zeta}^4\left(-\frac{1}{k_1^3k_2^3}-\frac{1}{k_2^3k_3^3}
-\frac{1}{k_3^3k_1^3}-\frac{2}{k_1^2k_2^2k_3^2}+\sum_{i\neq j \neq l}
\frac{1}{k_i k_j^2 k_l^3}\right),\\
& B^{\text{orth}.}_\zeta(k_1,k_2,k_3)=\frac{18}{5}f_{NL}^{\text{orth}.}
\mathcal{A}_{\zeta}^4\left(\!-\frac{3}{k_1^3k_2^3}\!-\!\frac{3}{k_2^3k_3^3}
\!-\!\frac{3}{k_3^3k_1^3}\!-\!\frac{8}{k_1^2k_2^2k_3^2}\!+3\!\!\sum_{i\neq j \neq l}
\frac{1}{k_i k_j^2 k_l^3}\!\right),
\end{align} 
\eseq
where in the last two expressions 
we for simplicity neglected the departures from scale invariance (the tilt).
The current observational constraints on $f_{NL}$ (at $68\%$ CL)
from the CMB measurements are \cite{Akrami:2019izv},
\be
f^{\text{loc}.}_{NL}=-0.9\pm 5.1\,,\qquad
f^{\text{eq}.}_{NL}=-26\pm 47\,,\qquad
f^{\text{orth}.}_{NL}=-38\pm 24\,.
\ee

The key objects for the description of LSS 
are the matter density contrast $\delta=(\rho-\bar \rho)/\bar\rho$ 
and velocity divergence $\Theta \propto \nabla \cdot {\bm u}$ 
fields. 
Here $\bar\rho$ is the average density of the Universe and 
${\bm u}$ is the peculiar flow velocity.
Initially small, these fields start growing after
recombination and become non-linear. This non-linear evolution
is captured by the cosmological
perturbation theory, which uses as a seed the {\em linear} fields 
$\delta_L$,
$\Theta_L$ evolved up to the present epoch as if the perturbations
were always in the linear regime.
In the case of adiabatic initial conditions corresponding 
to a growing mode the two linear fields
are identical. They are related to the curvature perturbation $\zeta$
by a transfer function $T(t_0;\k)$
which encodes the evolution
of perturbations from inflation, through recombination, up to the
present time
$t_0$. Thus we write,
\be
\label{eq:tf}
 \Theta_L (\k)= \delta_L(\k)\equiv T(t_0; \k) \zeta(\k)\,.
\ee
We will understand by {\em
  linear} statistics the properties of the field $\delta_L$.
Defining the linear matter power spectrum, 
\be 
\label{eq:initialPS}
\langle\delta_L(\k) \delta_L(\k')\rangle
=(2\pi)^3\delta^{(3)}_D(\k+\k')P_L(k)\;, 
\ee
we see that
\be
\label{earlyps}
 P_{L}(k)=[T(t_0;k)]^2P_{\zeta}(k)\;.
\ee
From this we immediately deduce 
the relation between the linear bispectrum of LSS and the primordial
one,
\be 
\begin{split}
  B_L(k_1,k_2,k_3) =  &  
  \sqrt{
  \frac{P_L(k_1)}{P_\zeta(k_1)}
  \frac{P_L(k_2)}{P_\zeta(k_2)}
  \frac{P_L(k_3)}{P_\zeta(k_3)}} 
  B_\zeta(k_1,k_2,k_3)\,.
  \label{earlybis}
  \end{split}
\ee
Note that the linear bispectrum 
is parametrically suppressed by the amplitude of the primordial
fluctuations. Indeed, neglecting the shape of the bispectrum one has,
schematically, 
\be
\label{eq:suppressfNL}
\frac{\langle \delta_L^3 \rangle}{(\langle \delta_L^2
  \rangle)^{3/2}}\sim \frac{B_L}{(P_L)^{3/2}}  
\sim f_{NL} \mathcal{A}_\zeta \ll 1\,.
\ee
The linear power spectrum (\ref{earlyps}) and bispectrum
(\ref{earlybis}) serve as the input for the equations
describing the non-linear evolution of LSS statistics at times after
recombination.

\subsection{Standard perturbation theory with non-Gaussian initial
  conditions}\label{sec:spt} 

We are interested in the correlation functions of
the overdensity  
and the velocity divergence fields,
whose time-evolution is governed by the 
following pressureless perfect fluid equations:
\bseq
\label{eq:hydro}
\begin{align}
 & \frac{\d\delta}{\d t} + \nabla \cdot [(1+\delta){\bm u}] = 0 \,, \\
 & \frac{\d{\bm u}}{\d t} + {\cal H}{\bm u} + ({\bm u}\cdot \nabla)
 {\bm u} =  -\nabla\phi \,, 
\end{align}
\eseq
where $\phi$ is the gravitational potential obeying the Poisson equation,
\be
\label{Poisson}
\nabla^2\phi=\frac{3}{2}{\cal H}^2\Omega_m\delta\,.
\ee
Here $t$ is the conformal time\footnote{Defined as $dt=d\tau/a(\tau)$, where $\tau$ is physical time and $a(\tau)$ is the scale factor of the Friedmann-Lemaitre-Robertson-Walker metric, \mbox{$ds^2=d\tau^2-a^2(\tau)dx_i dx^i$}.}, the spatial derivatives are taken with
respect to the comoving coordinates, 
${\cal H}=aH$ is the rescaled Hubble parameter, $a$ is the scale
factor and $\Omega_m$ is the matter
density fraction. 

The single-stream perfect fluid approximation breaks down at short scales due to
free-streaming of dark matter particles (also called `shell-crossing'). 
These effects show up when one computes loop contributions generated by hard modes. 
They are taken into account in the EFT of LSS, where the
influence of short scale nonlinearities on physics at large scales is
encapsulated by an effective stress tensor added to the r.h.s. of
(\ref{eq:hydro}) \cite{Baumann:2010tm,Carrasco:2012cv}. For
simplicity, we do not explicitly consider the EFT corrections in this paper, keeping in
mind that eventually they will have to be properly included.
The main goal of this paper is to study the non-linear effects of very long-wavelength (IR) modes, 
for which the perfect fluid description is sufficient. 

It is well-known \cite{Bernardeau:2001qr} that in
the case of an EdS universe dominated by non-relativistic matter
($\Omega_m=1$) 
the above equations can be cast
in a form free from any explicit time dependence. This is achieved by introducing  
the time parameter 
\be
\label{eta}
\eta=\ln D(t)\;,
\ee
where $D$ is the linear growth
factor\footnote{Following the standard practice, we normalize
  $D(t)$ to be equal to 1 at the present epoch.}, 
and appropriately rescaling the velocity divergence,
\be
\label{Thetaf}
\Theta = -\frac{\nabla \cdot {\bm u}}{{\cal H}f}~,\qquad\text{with}~
f=\frac{d\ln D}{d\ln a}\;.
\ee
For the realistic $\Lambda$CDM cosmology,
the above substitution leaves a small residual time dependence which,
however, has little effect on the dynamics \cite{Pietroni:2008jx}. 
Following conventional
practice we will neglect this explicit time dependence in the
equations, even though our analysis does not crucially depend on this restriction.

Within the above approximation Eqs.~\eqref{eq:hydro} can be rewritten
in Fourier space as\footnote{Our conventions are: 
$$\delta_\k\equiv \delta(\k)=\int d^3x\,\delta(\x) e^{-i\k\cdot \x}~,
\qquad\delta(\x)=\int \frac{d^3k}{(2\pi)^3}\,\delta_\k e^{i\k\cdot \x}\,.$$ 
Note that they differ from those used in 
\cite{Blas:2015qsi,Blas:2016sfa} by factors of $(2\pi)^3$.}
\bseq
\label{eq:eomsreal}
\begin{align}
& \d_\e\delta_\k -\T_\k=\int [dq]^2\,(2\pi)^3\delta^{(3)}_D(\k-\q_{12})
\alpha(\q_1,\q_2)\T_{\q_1}\delta_{\q_2}\,,
\label{eq:eomsreal1}\\
& \d_\e \Theta_\k +\frac{1}{2}\Theta_\k - \frac{3}{2}\delta_\k=  
\int
[dq]^2\,(2\pi)^3\delta^{(3)}_D(\k-\q_{12})\beta(\q_1,\q_2)\T_{\q_1}\T_{\q_2}\,,
\label{eq:eomsreal2}
\end{align}
\eseq
where we used the notations,
\[
[dq]^n\equiv \frac{d^3q_1...d^3q_n}{(2\pi)^{3n}}\,,\quad  \quad \q_{1...n}=\q_1+...+\q_n\,,
\]
and introduced the non-linear kernels
\be
\label{alphabetareal}
\alpha(\k_1,\k_2)\equiv\frac{(\k_1+\k_2)\cdot \k_1}{k_1^2}\,, \quad \quad 
\b(\k_1,\k_2)\equiv\frac{(\k_1+\k_2)^2(\k_1\cdot \k_2)}{2k_1^2k_2^2}\,. 
\ee
These equations can be solved perturbatively 
by using the following power series Ansatz:
\bseq
\label{SPTexp}
\begin{align}
& \delta_\k=\sum_{n=1}^\infty \big(D(\eta)\big)^n 
\int [dq]^n\, (2\pi)^3\delta^{(3)}_D(\k-\q_{1...n})F_n(\q_1,...,\q_n)
\delta_L(\q_1)...\delta_L(\q_n) \,,\\
& \T_\k=\sum_{n=1}^\infty \big(D(\eta)\big)^n 
\int [dq]^n\, (2\pi)^3\delta^{(3)}_D(\k-\q_{1...n})G_n(\q_1,...,\q_n)
\delta_L(\q_1)...\delta_L(\q_n) \,,
\end{align} 
\eseq
where $\delta_L$ is the linear density field 
and the non-linear kernels $F_n$ and $G_n$ are recursively derived
from \eqref{eq:eomsreal}. 
The correlation functions of the fields $\delta_\k$, $\Theta_\k$ are
computed by averaging the expressions (\ref{SPTexp}) using the
statistics of the linear field $\delta_L$. The resulting expressions
are conveniently represented as diagrammatic expansions in the number
of loops. 
This method is known as Eulerian standard perturbation theory (SPT)
\cite{Bernardeau:2001qr}. 
Typically, the linear statistics are assumed to be Gaussian and
adiabatic, i.e. they are fully characterized by the power spectrum
(\ref{eq:initialPS}). 
Recall that the initial conditions for structure formation
are set shortly after recombination when 
baryons and dark matter started to behave as a single pressureless
fluid, but the gravitational instability 
did not yet have enough time to form non-linear structures.
For simplicity, we neglect non-Gaussianity generated by the second order effects at radiation
domination and recombination 
\cite{Fitzpatrick:2009ci,Huang:2013qua,Tram:2016cpy}.

Suppose now that in addition to the power spectrum
the linear distribution of the density contrast
is also characterized by a non-trivial 3-point function,
\be
\begin{split}
  & \corr{\delta_L(\k_1)\delta_L(\k_2)\delta_L(\k_3)}  
= (2\pi)^3 \delta^{(3)}_D(\k_1+\k_2+\k_3)B_{L}(k_1,k_2,k_3)\,.
  \label{eq:bafterrec}
  \end{split}
\ee
This generates additional terms in the SPT perturbative expansion,
which originate from 
averages of 
odd number of fields. For instance, 
the leading-order (tree-level) matter 
bispectrum and the
1-loop correction
to the matter power spectrum read,
\bseq
\begin{align}
&B_{\rm SPT}^{\rm tree}(\eta;k_1,k_2,k_3)=
2\big(D(\eta)\big)^4
\sum_{1\leq i<j\leq 3}F_2(\k_i,\k_j)P_L(k_i)P_L(k_j)
+\big(D(\eta)\big)^3 B_L(k_1,k_2,k_3)\,,\\
&P_{\text{SPT}}^{\text{1-loop}}(\e;k)=\big(D(\eta)\big)^4
\bigg(2\int [dq] P_{L}(q)P_{L}(|\k-\q|)\big(F_2(\q,\k-\q)\big)^2\notag\\ 
&\qquad+6 P_{L}(k)\int [dq] P_{L}(q) F_3(\k,\q,-\q)\bigg)
+ 2 \big(D(\eta)\big)^3 
\int [dq] B_{L}(q,|\k-\q|,k) F_2(\q,\k-\q)\,.
\label{eq:spt1loop}
\end{align}
\eseq
Note that the primordial non-Gaussian contributions have one less
power of the growth factor $D(\eta)$ compared to the terms
coming from the non-linear evolution.

Let us discuss an important point. The kernels $\a$ and $\b$ entering
the Euler equations have poles if one of their arguments vanishes, see
Eqs.~(\ref{alphabetareal}). These IR singularities are inherited by
the non-linear kernels $F_n$, $G_n$; for example,
\[
F_2(\k,\q)\Big|_{|\q|\to 0}\simeq \frac{\k\cdot\q}{2q^2}\;.
\]
This leads, in turn, to IR poles in the integrands of the loop
expressions. The origin of the IR singularities can be traced back to
large displacements of short-wavelength density perturbations by
long-wavelength bulk flows. This effect should, however, cancel in the
correlation functions of fields taken at the same moment of
time. Indeed, the cancellation of IR singularities in equal-time
correlators is a well-known property in Gaussian SPT, see \cite{Blas:2013bpa} and
references therein. For non-Gaussian initial conditions, the
cancellation of contributions due to the IR poles can be tracked
explicitly at the lowest orders of the perturbation theory; however,
we are not aware of a general proof in the literature. Clearly, it is
desirable to have a framework where spurious IR singularities are
absent altogether. Such framework is provided by the time-sliced
perturbation theory \cite{Blas:2015qsi} described in the
subsequent sections. 

It is important to stress that the IR
poles discussed above should be distinguished from those that
may be present in the primordial bispectrum itself and thus are
physical. Throughout the paper
`IR poles' or `IR singularities' will only refer to the 
singularities originated from the non-linear mode coupling.

\section{Review of  Gaussian TSPT}\label{sec:tspt}

The main idea of the TSPT approach
is to substitute the time evolution of the overdensity and velocity
divergence fields, $\delta$ and $\Theta$, by that of the their time
dependent probability distribution functional (PDF).
This idea is natural when one is only interested in equal time
correlation functions. 
For adiabatic initial conditions only one of the two fields is
statistically independent. 
It is convenient to choose the velocity divergence field $\T$ as an
independent variable  
and its PDF will be denoted by
$\mathcal{P}[\T;\e]$.
At any moment in time, the field $\delta$ can be expressed in terms of  $\T$
as 
\be
\label{eq:psi1}
\delta(\k)\equiv\delta[\T;\e,\k]= \sum_{n=1}^\infty\frac{1}{n!}
\int [dq]^n (2\pi)^3\delta^{(3)}_D\big(\k-
\q_{1\ldots n}\big)\,K_n(\e;\q_1,...,\q_n)
\prod_{j=1}^n\T(\q_j)  \,.
\ee
Note that, in contrast to the SPT formulae (\ref{SPTexp}), we have here on the
r.h.s. the full non-linear field $\Theta(\q)$. Recursion relations for
the kernels $K_n$ are given in Appendix~\ref{app:tsptvert}.
Equation \eqref{eq:psi1} can be used to eliminate the density field
from Eq.~\eqref{eq:eomsreal2} and obtain the 
following equation for the velocity divergence,
\be
 \label{eq:In}
\dot\T(\k)= \mathcal{I}[\T;\eta]\equiv \sum_{n=1}^\infty\frac{1}{n!}
\int [dq]^n (2\pi)^3\delta^{(3)}_D\big(\k-\q_{1\ldots n}\big)\,I_n(\e;\q_1,...,\q_n)
\prod_{j=1}^n\T(\q_j)  \,,
\ee
with $I_1\equiv 1$ corresponding to the adiabatic mode
in the perfect fluid approximation. The other kernels $I_n$
can be derived from the fluid equations \eqref{eq:eomsreal}, see
Appendix~\ref{app:tsptvert}. In the EdS approximation both kernels
$K_n$ and $I_n$ are time independent.

One introduces the generating functional,
\be
\label{eq:ztfp}
Z[J,J_\delta;\e]=\int [\mathcal{D}\T]\;{\mathcal P}[\T;\e]\;
\exp\bigg\{\int [dk] \T(\k) J(-\k)+\int[dk]
\delta[\T;\e,\k]J_{\delta}(-\k)\bigg\}\,. 
\ee
Equal-time correlation functions for $\T$ and $\delta$ are obtained by
taking functional derivatives of $Z$ with respect to the external sources $J$ or
$J_\delta$, respectively.
For example, the matter power spectrum is given by
\be
 (2\pi)^3\delta^{(3)}_D(\k+\k')\, P(\e;k) = 
\frac{\delta^2 Z}{\delta J_\delta(-\k)\delta J_\delta(-\k')}\bigg|_{J=J_\delta=0}\,.
\ee
The conservation of probability implies the Liouville equation for the
probability density 
functional, 
\be
\frac{\d}{\d \e} \mathcal{P}[\T;\e] + 
\int [dk]\frac{\delta}{\delta \Theta(\k)}
\big(\mathcal{I}[\T;\e]\mathcal{P}[\T;\e]\big)=0\,.
\ee
It is useful to expand $\ln\mathcal{P}[\T;\e]$ as a power series 
in $\T$,
\be
\label{eq:statweight}
\mathcal{P}[\T;\e]=\mathcal{N}^{-1}\exp\Bigg\{-\sum_{n=1}^{\infty}\frac{1}{n!}\int [dk]^n\;
\G_n^{tot}(\e;\k_{1},...,\k_{n})\; \prod^n_{j=1} \T(\k_j)\Bigg\} \,,
\ee
where ${\cal N}$ is a normalization factor. Then, using 
Eq.~\eqref{eq:In} we obtain the following chain of equations 
for the vertices $\G^{tot}_n$,
\begin{equation}
\begin{split}
  & \dot{\Gamma}_n^{tot}(\eta;\narg) + \sum_{m=1}^n 
 \sumsig \frac{I_m(\eta;\ks{1},...,\ks{m})}{m!(n-m)!}\,
  \Gamma_{n-m+1}^{tot}\bigg(\eta; \sum_{i=1}^m \ks{i},\ks{m+1},...,\ks{n}\bigg) \\
   &=  (2\pi)^3\dirdel_D\left(\k_{1\ldots n} \right) \int [dq]\, 
I_{n+1}(\eta;\q,\k_1,...,\k_n)\;,
  \label{eq:liouvillegamma}
\end{split}
\end{equation}
where the sum in the second term on the l.h.s. is taken over all
permutations $\sigma$ of $n$ indices.
It is convenient to decompose the solution of this equation into two pieces,
\be 
\G_n^{tot}=\G_n+C_n\,,
\ee
where $\G_n$ is the solution of Eq.~\eqref{eq:liouvillegamma} with
vanishing r.h.s. and with the initial conditions reflecting 
the initial statistical distribution,
whereas $C_n$ is the solution of the inhomogeneous equation with
vanishing initial conditions. 
The vertices $\G_n$ thus
have a physical meaning of 1-particle irreducible contributions to the
tree-level correlators with amputated external propagators; the
vanishing of the average velocity dispersion, $\langle
\Theta(\k)\rangle=0$, implies $\G_1=0$.  
On the other hand,
$C_n$ are {\it counterterms} that cancel certain ultraviolet divergences
appearing in the loop corrections \cite{Blas:2015qsi}. 

The vertices $\G_n$ and $C_n$ satisfy a hierarchy
of equations which replaces the dynamical equations of SPT.  
From now on we specify to the EdS approximation where the kernels $I_n$ are
constant in time. 
In order to find $\G_n$, we use an Ansatz which separates time and
momentum dependence, 
\be 
\G_n(\eta;\k_1,...,\k_n)=\sum_{l=2}^n
e^{-l\eta}\,\G_n^{(l)}(\k_1,...,\k_n)~,~~~~~~
n\geq 2\;.
\ee
This leads to the following recursion relations for $\G_n^{(l)}$ with
$l<n$, 
\begin{equation}
\begin{split}
  \gxx{l}{n}(\narg)\!=-\frac{1}{n-l}\!\!\sum_{m=2}^{n-l+1} \!\sumsig
  \frac{I_m(\ks{1},...,\ks{m})}{m!(n-m)!}  
  \gxx{l}{n-m+1}\!\left(\sum_{i=1}^m \ks{i},\ks{m+1},...,\ks{n}\!\right)\!.
  \label{eq:gl}
\end{split}
\end{equation}
Note that the r.h.s. contains only vertices $\G^{(l)}_{n'}$ with $n'$
less than $n$, implying that all $\G_n^{(l)}$, $l<n$, are uniquely
determined from vertices of lower orders. The remaining function
$\G_n^{(n)}$ must be fixed from the initial conditions. 
Not to overload the paper with unnecessary formulae, we 
do not reproduce here the equations for the counterterms
$C_n$ which can be
found in \cite{Blas:2015qsi}. The upshot is that $C_n$ are completely
fixed in terms of the kernels $I_n$ and are time independent. 

To sum up, the PDF (\ref{eq:statweight}) is fully
specified by providing the `diagonal' vertices $\G_n^{(n)}$, $n\geq
2$. These define the early-time asymptotics of the full vertex functions,
\be 
\label{eq:defbar}
\lim_{\e_i\to -\infty}\; e^{n\eta_i}\,\G_n(\eta_i;\narg)=\gxx{n}{n}(\narg)\,.
\ee 
Note that in the last formula we used a convenient trick of sending 
the initial time $\eta_i$ to $-\infty$ assuming the validity of the pressureless fluid description (see Eqs.~\eqref{eq:hydro})
at all times. Of course, this does not correspond to a realistic 
Universe dominated by radiation at an early epoch. 
Physically, the initial data for Eq.~\eqref{eq:liouvillegamma} should be set somewhen 
after recombination. 
However, mapping the initial conditions to $\eta_i=-\infty$ by formally  
extrapolating the EdS description towards the past  
leads to a great simplification of formulas.


For the Gaussian initial conditions the solution to \eqref{eq:gl} is
greatly simplified.  
In that case the PDF ${\cal P}[\T;\e]$ should reduce
to a Gaussian weight at early times, upon rescaling the field with the
linear growth factor,
$\T\mapsto\tilde\T=\T/D(\eta)$. In other words, one requires,
\be
\label{calPlim}
\lim_{\eta\to-\infty}{\cal P}\big[e^\e\tilde\T;\e\big]=
{\cal N}^{-1}\exp\bigg\{-\int [dk]\frac{\tilde\T_\k\tilde\T_{-\k}}{2 P_L(k)}\bigg\},
\ee
where $P_L(k)$ is the linear power spectrum. This implies the
following initial conditions for the vertices,
\bseq
\label{eq:initGauss}
\begin{align}
&\Gamma_2^{(2)}(\k_1,\k_2)
=\frac{(2\pi)^3\delta^{(3)}_D(\k_1+\k_2)}{P_L(k_1)}\,,
\label{eq:initial}\\
& \Gamma_n^{(n)}(\narg) =0\,,~~~~~n>2\;.
\end{align}
\eseq
Then all $\G^{(l)}_n$ with $l>2$ vanish, leaving the solution, 
\bseq 
\label{eq:gaussgn}
\begin{align}
\label{treeact}
& \G_n(\narg)=\frac{1}{g^2(\e)}\bar\G_n(\narg)\,,\\
&\ggv{n}(\narg)
=-\frac{1}{n-2}\sum_{m=2}^{n-1}\sumsig
\frac{I_m(\ks{1},...,\ks{m})}{m!(n-m)!} 
\ggv{n-m+1}\bigg(\sum_{i=1}^m \ks{i},\ks{m+1},..., \ks{n}\!\bigg),
\label{Gaussrecurs}
\end{align}
\eseq
where we have introduced the notations $\bar\G_n\equiv \G_n^{(2)}$ and 
\be\label{coupling}
  g(\eta) \equiv e^\eta = D(\e)\,.
\ee
Remarkably, in the case of 
Gaussian initial conditions and the EdS background, all 
$\G_n$ vertices have universal
dependence on time through the factor $g^{-2}(\e)$.
As will be discussed shortly, $g(\e)$ plays the role of expansion
parameter in TSPT.  
Due to momentum conservation, the vertices are proportional to a 
$\delta$-function of the sum of their arguments.
In what follows we use primes to denote the
quantities stripped off such $\delta$-functions, 
\be
\label{eq:tildeGC}
\bar \G_n(\narg)=(2\pi)^3\delta^{(3)}_D(\k_{1\ldots n})
\,{\bar \G}'_n(\k_1,...,\k_n)
\,.
\ee 
Note that the recursion relations (\ref{Gaussrecurs}) and the initial
conditions (\ref{eq:initial}) imply that all $\bar\G_n$ vertices are
functionals of the linear power spectrum $P_L(k)$.

\begin{figure}
\begin{align*}
\begin{fmffile}{example-ps}
\parbox{90pt}{
\begin{fmfgraph*}(80,80)
\fmfpen{thick}
\fmfleft{l1}
\fmfright{r1}
\fmf{plain,label=${\bf k}$}{l1,r1}
\end{fmfgraph*}}
\end{fmffile}&=~g^2(\e) P_L(k), ~~
\begin{fmffile}{example-bisp_f}
\parbox{100pt}{
\begin{fmfgraph*}(90,70)
\fmfpen{thick}
\fmfleft{l1,l2}
\fmfright{r1}
\fmf{plain,label=${\bf k}_1$,label.side=right}{l1,b1}
\fmf{plain,label=${\bf k}_2$,label.side=left}{l2,b1}
\fmf{plain,label=${\bf k}_3$}{b1,r1}     
\end{fmfgraph*}}
\end{fmffile}=-\frac{1}{g^{2}(\e)}\frac{1}{3!}\bar\G_3(\k_1,\k_2,\k_3)\\
\begin{fmffile}{example-C1}
\parbox{80pt}{
\begin{fmfgraph*}(75,75)
\fmfpen{thick}
\fmfleft{l1}
\fmfright{r1}
\fmf{plain,label=${\bf k}$}{l1,r1}
\fmfv{d.sh=cross,d.si=.15w}{l1}
\end{fmfgraph*}}
\end{fmffile}&=-C_1(\k),~~~~~
\begin{fmffile}{composite}
\parbox{80pt}{
\begin{fmfgraph*}(75,75)
\fmfpen{thick}
\fmfleft{l1}
\fmfright{r5,r6}
\fmfv{d.sh=circle,d.filled=full,d.si=3thick}{b1}
\fmf{phantom_arrow,tension=4,label=$\k$,l.s=left}{l1,b1}
\fmf{plain}{b1,v6}
\fmf{plain,label=$\q_1$,l.s=left}{v6,r6}
\fmf{plain}{b1,v5}
\fmf{plain,label=$\q_2$,l.s=left,l.d=4}{v5,r5}
\end{fmfgraph*}}
\end{fmffile}=\frac{1}{2}(2\pi)^3\delta^{(3)}_D(\k-\q_{12})
\,K_2(\q_1,\q_2)
\end{align*}
\caption{Example of TSPT Feynman rules.\label{fig:feynmanrules}}
\end{figure}
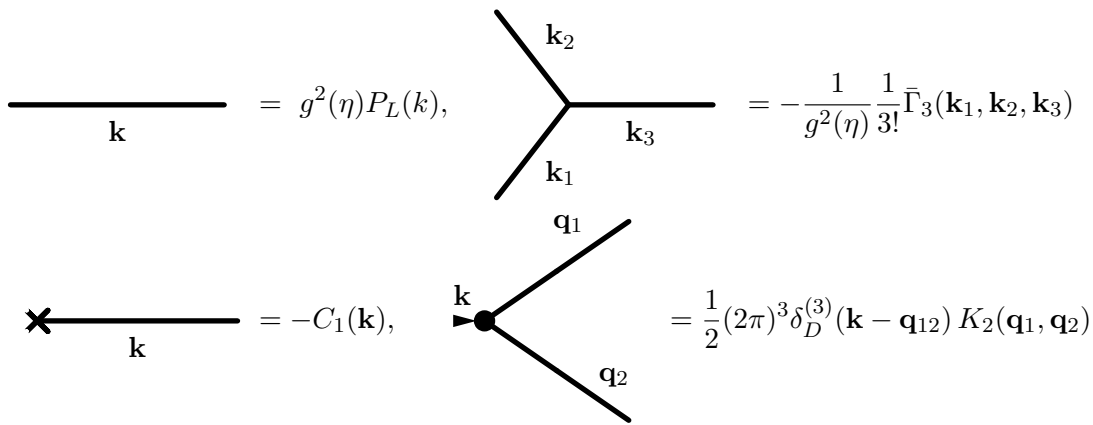

The TSPT perturbative series is obtained upon expanding the
generating functional \eqref{eq:ztfp} over the Gaussian part of
$\mathcal{P}$. Following the standard rules of quantum field theory,
this expansion can be represented as a sum of Feynman diagrams. These
are built from vertices corresponding to $\G_n$, $n\geq 3$, and lines
corresponding to the `propagator' $g^2 P_L(k)$, see
Fig.~\ref{fig:feynmanrules}. One should also include vertices
corresponding to counterterms $C_n$, $n\geq 1$, in order to subtract
certain ultraviolet divergences of loop diagrams.
To compute an $n$-point correlation
function of the velocity divergence $\T$
one needs to draw all diagrams with $n$ external legs. It is
straightforward to see that diagrams with higher number of loops are
proportional to higher powers of $g(\e)$. Hence, $g(\e)$ should be
interpreted as the coupling constant of the theory. 
For the correlators of the density field $\delta$ one 
uses the expression (\ref{eq:psi1}) which is reminiscent of
expressions for composite operators in quantum field theory. It
gives rise to additional
vertices proportional to the kernels $K_n$; these are denoted by an
external arrow.

\section{TSPT with primordial non-Gaussianity and Feynman rules}
\label{sec:tsptPNG}

In the presence of primordial non-Gaussianity the initial conditions
(\ref{eq:initGauss}) get modified. One expects now a non-vanishing
initial value of the 3-point vertex, $\G_3^{(3)}\neq 0$. The
early-time asymptotics of the PDF become,
\be 
\label{eq:pdfini}
\lim_{\eta\to -\infty}
\mathcal{P}[e^\eta\tilde\T;\e]={\cal N}^{-1}\exp\bigg\{-\int [dk]
\frac{\tilde\T_{\k}\tilde\T_{-\k}}{2 P_{L}(k)} 
-\int\frac{[dk]^3}{3!} 
\G^{(3)}_3(\k_1,\k_2,\k_3)\, \tilde\T_{\k_1}\tilde\T_{\k_2}\tilde\T_{\k_3}
\bigg\}\,.
\ee
Substituting this expression into the generating functional
(\ref{eq:ztfp}) and taking variational derivatives with respect to the
external sources one derives the correlation functions of the
$\T$-field at early times. The latter are to be matched to the linear
statistics. Assuming that the non-Gaussian contribution is small, as
will be confirmed shortly, we can restrict to linear order in the
cubic vertex $\G_3^{(3)}$ and obtain the matching condition,
\be
\label{NGmatching}
(2\pi)^3\delta^{(3)}_D(\k_{123})\,B_{L}(k_1,k_2,k_3)=
 -\G_3^{(3)}(\k_1,\k_2,\k_3) P_{L}(k_1)P_{L}(k_2)P_{L}(k_3)\,.
\ee
We observe that $\G_3^{(3)}$ is proportional to the linear
bispectrum. Let us estimate the size of the cubic term in
(\ref{eq:pdfini}). Taking for the characteristic amplitude of the
modes $\tilde\T_\k\sim\sqrt{P_L(k)}$ we obtain 
${\G_3^{(3)}}'\tilde\T^3\sim B_L/(P_L)^{3/2}$. As discussed before,
the latter quantity is of order $f_{NL}{\cal A}_\zeta$ which is much
smaller than unity, given the existing bounds on non-Gaussianity. This
justifies our expansion to linear order in $\G_3^{(3)}$.

The TSPT vertices receive a contribution seeded by the primordial
non-Gaussianity,
\be 
\label{eq:nongaussgn}
\G_n(\eta;\narg)=\frac{1}{g^2(\eta)}\bar\G_n(\narg)
+\frac{1}{g^3(\eta)}{\bar\G}^{NG}_n(\narg)\,,
\ee
where 
\bseq 
\label{eq:g3ng}
\begin{align}
\label{G3NG}
& \bar\G_3^{NG} (\k_1,\k_2,\k_3)\equiv\G_3^{(3)} (\k_1,\k_2,\k_3)  = -(2\pi)^3\delta_D^{(3)}(\k_{123})\,
\frac{B_L(k_1,k_2,k_3)}{P_L(k_1)P_L(k_2)P_L(k_3)}\;,\\
\label{GNNG}
&\bar\G_n^{NG}(\narg)\!=\!-\frac{1}{n-3}\!\sum_{m=2}^{n-2}\sum_{\sigma}
\frac{I_m(\ks{1},...,\ks{m})}{m!(n-m)!} 
  \bar\G_{n-m+1}^{NG}\bigg(\!\sum_{i=1}^m \ks{i},\ks{m+1},...,\ks{n}\!\bigg).
\end{align}
\eseq
Similarly to the Gaussian vertices $\bar\G_n$, the non-Gaussian terms
$\bar\G_n^{NG}$ can be systematically constructed by induction. An
explicit expression for $\bar\G_4^{NG}$ is given in
Appendix~\ref{app:1loopps}. 

A comment is in order. The non-Gaussian vertices are proportional to
an extra inverse power of the coupling $g$ compared to the Gaussian
part. This may give a wrong impression that they are
enhanced. Actually, the contrary is true: they are suppressed due to
the small amplitude of the primordial non-Gaussianity. Indeed, let us
estimate the ratio between the non-Gaussian and Gaussian vertices. As
usual, we neglect the momentum dependence and focus only on the
overall amplitude. We have, 
\be
\label{eq:G3overG2}
\frac{\bar\G_n^{NG}}{g\,\bar\G_n}\sim 
\frac{1}{g}\cdot\frac{B_L}{P_L^3} \cdot P_L
\sim f_{NL}\frac{{\cal A}_\zeta}{g\sqrt{P_L}}\;,
\ee
where we have used that $\bar\G_n^{NG}$ and $\bar\G_n$ are
proportional to $\bar\G_3^{NG}$ and $\bar\G_2$ given by
Eqs.~(\ref{G3NG}), (\ref{eq:initial}) respectively. In the last
expression we also used the estimate (\ref{eq:suppressfNL}). The ratio
$g(\eta)\sqrt{P_L}/{\cal A}_\zeta$ is nothing, but the transfer
function $T(\eta)$, which is much bigger than unity at all relevant
times and momenta. So, unless $f_{NL}$ is very large, the r.h.s. in
(\ref{eq:G3overG2}) is much smaller than one. This suggests that the
primordial non-Gaussianity should be treated perturbatively, on top of
the standard vertices generated by non-linear clustering.
It is convenient to consider $f_{NL}$ as a bookkeeping parameter 
which in all 
expressions is associated to the suppression (\ref{eq:G3overG2}). 
In other words, in the case of primordial non-Gaussianity 
we should also make an expansion in $f_{NL}$
on top of the TSPT expansion in powers of $g(\eta)$.
For practical purposes, it is enough 
to keep terms up to linear order in~$f_{NL}$. 

We now discuss the diagrammatic technique. In principle, one could
just modify the explicit structure of vertices $\G_n$ and keep the
same Feynman rules as in the Gaussian case. 
The propagator is still determined by the Gaussian part and is given by
the linear power spectrum. 
The other building blocks of TSPT, 
such as the counterterms $C_n$ and kernels $K_n$, are defined by the dynamical
equations of motion ($I_n$ kernels) and thus are not modified when
including non-Gaussianity. 
However, since the non-Gaussian vertices are suppressed as \eqref{eq:G3overG2} and 
have a different scaling with the TSPT coupling constant,  
it is useful to single them out by introducing new graphical
notations. We will denote them with an open star, e.g. for the
non-Gaussian 3-point vertex we will use:
\begin{equation}
\begin{fmffile}{example-bisp_ng}
\parbox{100pt}{
\begin{fmfgraph*}(90,70)
\fmfpen{thick}
\fmfleft{l1,l2}
\fmfright{r1}
\fmf{plain,label=${\bf k}_1$,label.side=right}{l1,b1}
\fmf{plain,label=${\bf k}_2$,label.side=left}{l2,b1}
\fmf{plain,label=${\bf k}_3$}{b1,r1}     
\fmfv{decor.shape=pentagram,decor.filled=empty,decor.size=10thick,label=$ $,l.a=120,l.d=.07w}{b1}
\end{fmfgraph*}}
\end{fmffile}
~~= 
  ~-\frac{1}{g^3(\e)}\,\frac{1}{3!}\,\bar{\Gamma}_3^{NG}(\k_1,\k_2,\k_3)\;.
  \label{eq:ngvertexrule}
\end{equation} 
Each non-Gaussian vertex contains one power of $f_{NL}$, 
thus in the diagrammatic expansion we should keep only the graphs containing at most one  
non-Gaussian element.
For instance,
the Feynman diagrams contributing to the 1-loop non-Gaussian matter power spectrum 
are given by 
\be
\label{eq:Ddelta}
P^{NG,\,\text{1-loop}}_{\delta\delta}(\eta;k)=  
\quad
\begin{fmffile}{4-vertex-ex1}
\parbox{80pt}{
\begin{fmfgraph*}(80,35)
\fmfpen{thick}
\fmfkeep{1loop}
\fmfleft{l1,l2}
\fmfright{r1,r2}
\fmf{plain,label=$ $,l.s=right}{l1,b1}
\fmf{plain,label=$ $,l.s=right}{b1,r1}
\fmf{phantom}{l2,u1}
\fmf{phantom}{u2,r2}
\fmf{plain,left=0.5,tension=0.01,l.side=left}{b1,u1}
\fmf{plain,left=0.5,tension=2,label=$ $}{u1,u2}
\fmf{plain,right=0.5,tension=0.01}{b1,u2}
\fmfv{decor.shape=pentagram,decor.filled=empty,decor.size=7thick,label=$ $,l.a=120,l.d=.07w}{b1}
\end{fmfgraph*}}
\end{fmffile}
~
+
\quad
\begin{fmffile}{3-vertex-ex1}
\begin{fmfgraph*}(80,1)
\fmfpen{thick}
\fmfkeep{1loop_2}
\fmfleft{l1}
\fmfright{r1}
\fmf{plain,label=$ $}{l1,v1}
\fmf{plain,label=$ $}{v2,r1}
\fmf{plain,left=1.0,tension=0.5,label=$ $}{v1,v2}
\fmf{plain,left=1.0,tension=0.5,label=$ $,l.side=left}{v2,v1}
\fmfv{decor.shape=pentagram,decor.filled=empty,decor.size=7thick,label=$ $,l.a=120,l.d=.07w}{v1}
\fmfposition
\end{fmfgraph*}
\end{fmffile}
\quad 
+	
~~	   	    
\begin{fmffile}{cs3toy}
\parbox{80pt}{
\begin{fmfgraph*}(70,60)
\fmfpen{thick}
\fmfkeep{1loop_2}
\fmfleft{l1}
\fmfright{r1}
\fmfv{d.sh=circle,d.filled=full,d.si=.01w,label=$ $,l.a=120,l.d=.05w}{v1}
\fmf{phantom_arrow,tension=2,label=$ $,l.s=right}{l1,v1}
\fmf{plain,tension=2,label=$ $,l.s=right}{v2,r1}
\fmf{plain,left=1,tension=0.5,label=$ $}{v1,v2}
\fmf{plain,left=1,tension=0.5,label=$ $,l.side=left}{v2,v1}
\fmfv{decor.shape=pentagram,decor.filled=empty,decor.size=7thick,label=$ $,l.a=120,l.d=.07w}{v2}
\fmfposition
\end{fmfgraph*}}
\end{fmffile} 
\ee
One can show that, similarly to the Gaussian case, the TSPT result for
non-Gaussian contributions at linear order
in $f_{NL}$ and a given order in $g$ coincides with the result of SPT at
the corresponding loop order. We verify this explicitly for the
one-loop corrections to the power spectrum (\ref{eq:Ddelta}) 
in Appendix~\ref{app:1loopps}; verification for the
one-loop bispectrum was performed in \cite{thesis}.

At first glance, the new diagrammatic technique may seem somewhat excessive, 
as the calculation of the non-Gaussian contribution to the power spectrum in TSPT
involves three graphs instead of only one in SPT. However, the new
representation has an important advantage: 
as will be shown in the next section, all TSPT graphs are
explicitly IR safe. 
This property allows, in particular, 
for a resummation of physical IR contributions, which are enhanced if the 
initial statistics have oscillating features. 

It is straightforward to generalize the non-Gaussian TSPT formalism to
include the redshift space distortions along the lines of 
Ref.~\cite{Ivanov:2018gjr}.
To this end, one describes the mapping from the real to redshift space
by means of a fictitious free motion of matter particles along the
$z$-direction. This evolution proceeds in fictitious 
`redshift time' ${\cal F}$ that
varies from $0$ to $f$, where $f$ is the logarithmic growth rate
defined in (\ref{Thetaf}) (not to be confused with $f_{NL}$). The
fictitious flow transforms the PDF of the cosmological
fields. Applying the TSPT technique to 
this flow produces a set of equations for the redshift space vertices,
which we will distinguish by the superscript `(s)'. These equations
are linear and therefore do not mix the Gaussian and non-Gaussian
parts of the PDF. Thus, we can directly use the formulas from
\cite{Ivanov:2018gjr} and write,
\be
\label{eqredshift}
\frac{\d}{\d{\cal F}}\bar\G_n^{({\rm s})NG}({\cal F};\narg)
+\sum_{i<j}^n I_2^{({\rm s})}(\k_i,\k_j)
\bar\G_{n-1}^{({\rm s})NG}({\cal
  F};\k_1,...,\check\k_i,...,\check\k_j,...,\k_i+\k_j)=0\;,
\ee  
where
\be
\label{RSDkernel}
I^{({\rm s})}_2(\k_1,\k_2)= \frac{(\k_1+\k_2)^2 k_{1,z}k_{2,z}}{k_1^2k_2^2}
\ee
is the redshift space kernel and the notation $\check\k_i$ means that
the momentum $\k_i$ is absent from the arguments of the corresponding
function. These equations should be integrated from ${\cal F}=0$
with the `initial conditions' set by the TSPT vertices in real space,
\be
\label{RSDinit}
\bar\G_n^{({\rm s})NG}(0;\narg)=\bar\G_n^{NG}(\narg)\;.
\ee
The final redshift space vertices are read off at ${\cal F}=f$ and
have the form of polynomial expressions in the logarithmic growth rate
$f$; explicitly, the lowest vertices are,
\bseq
\begin{align}
&\bar\G_3^{({\rm s})NG}(\k_1,\k_2,\k_3)=\bar\G_3^{NG}(\k_1,\k_2,\k_3)\;,\\
&\G^{({\rm s})NG}_{4}(\k_1,\k_2,\k_3,\k_4)
\!=\!{\bar \G}^{NG}_{4}(\k_1,\k_2,\k_3,\k_4)\!-\!f\sum_{i<j}^{4}
I_2^{(s)}(\k_i,\k_j){\bar \G}^{NG}_{3}(\k_i+\k_j,\k_l,\k_m)
\Big|_{\begin{smallmatrix}
l< m\\l,m\neq i,j
\end{smallmatrix}}\,.
\end{align}
\eseq
The expressions for the other building blocks of TSPT in redshift
space, such as the Gaussian vertices $\bar\G_n^{({\rm s})}$,
counterterms $C_n^{({\rm s})}$ and kernels $K_n^{({\rm s})}$, are not
modified by the primordial non-Gaussianity and can be found
in~\cite{Ivanov:2018gjr}. 

Before closing this section, let us briefly discuss the case of an
arbitrary primordial non-Gaussianity encoded by 1-particle irreducible primordial
$l$-point functions,
\be
\label{priml}
\langle\zeta(\k_1)\ldots\zeta(\k_l)\rangle\Big|_\text{1-PI}
=(2\pi)^3\delta_D^{(3)}(\k_{1\ldots l})\,
G^{(l)}_{\zeta}(\k_1,\ldots,\k_l)\;.
\ee
These give rise to linear $l$-point functions,
\be
\label{linl}
G^{(l)}_{L}(\k_1,\ldots,\k_l)=\prod_{i=1}^l
\sqrt\frac{P_L(k_i)}{P_\zeta(k_i)}\,G^{(l)}_{\zeta}(\k_1,\ldots,\k_l)\;.
\ee
The TSPT vertices are then found from the recursion relations
(\ref{eq:gl}) where now the initial conditions for the diagonal
elements $\G_l^{(l)}$ are non-zero. Assuming that the linear
non-Gaussianity can be treated perturbatively, one obtains,
\be 
\label{eq:incondNGgen}
\Gamma_l^{(l)}(\k_{1},...,\k_{l}) =
-(2\pi)^3\delta_D^{(3)}(\k_{1\ldots l})
\frac{G^{(l)}_{L}(\k_1,...,\k_l)}{P_L(k_1)\ldots P_L(k_l)}\;.
\ee
The full TSPT vertices can be written in the form,
\be
\label{eq:nongaussgn2}
\G_n(\eta;\narg)=\frac{1}{g^2(\eta)}\bar\G_n(\narg)
+\sum_{l=3}^n \frac{1}{g^l(\eta)}{\G}^{(l)}_n(\narg)\,,
\ee
where the $l$-th term in the sum is generated from (\ref{eq:incondNGgen})
through the chain equations (\ref{eq:gl}). Note that a primordial
$l$-point function contributes only into the TSPT vertices with 
$n\geq l$.
Similarly to the bispectrum case, one can introduce new diagrammatic 
notations for the 
non-Gaussian vertices with different $l$, whose magnitude will be
controlled by 
\be
\label{lNGestim}
f_{NL}^{(l)} \bigg(\frac{{\cal A}_\zeta}{g\sqrt{P_L}}\bigg)^{l-2},
\ee
where $f_{NL}^{(l)}$ stands for the overall normalization of the
primordial $l$-point function.

\section{IR safety}
\label{sec:ir}

The TSPT framework is naturally adapted to produce the equal-time
correlation functions of the cosmological fields. The latter are
protected from IR singularities by the
equivalence principle 
\cite{Scoccimarro:1995if,Kehagias:2013yd,Peloso:2013zw,
Creminelli:2013mca,Horn:2014rta}.
Thus, one expects the TSPT formalism to be manifestly IR safe. Indeed,
it was shown in 
Ref.~\cite{Blas:2015qsi} 
that in the Gaussian TSPT the kernels $K_n$, counterterms $C_n$ and
vertices $\bar\G_n$ are free from IR singularities. We now extend the
proof to the non-Gaussian case. As $K_n,C_n,\bar\G_n$ are not affected
by the presence of non-Gaussianity, we need to analyze only the new
non-Gaussian vertices $\bar\G_n^{NG}$.

We are interested only in the IR singularities
that may appear due to the dynamics of gravitational clustering and 
do not consider the poles that can be present
in  
the initial bispectrum itself\footnote{This is for example the case
  for local primordial non-Gaussianity.}.
Thus, for the sake of the argument, we will assume that the linear
bispectrum is free from singularities. Then the seed
non-Gaussian vertex $\bar\G_3^{NG}$ is IR safe, see Eq.~(\ref{G3NG}). 
In order to show IR safety for an arbitrary vertex, we proceed by induction.
Consider a non-Gaussian $(n-1)$-point vertex whose momentum arguments
can be either soft (denoted by $q$) 
or hard (denoted by $k$) satisfying 
\be
q\ll k\,. 
\ee
Suppose that $\bar\G^{NG}_{n-1}$ is IR safe when any subset of its
arguments uniformly go to zero,   
\be
\q_j=\varepsilon \, \tilde{\q}_j \,,\quad \quad \varepsilon\to 0\,,\quad\tilde{\q}_j-\text{fixed}\,.
\ee
Consider now the recursion relation (\ref{GNNG}) determining the
vertex $\bar\G_n^{NG}$. According to the formulas from the
Appendix~\ref{app:tsptvert}, the kernels $I_{n'}$ with $n'\geq 3$ are
proportional to $K_{n'}$ and hence are IR safe. Further, the IR
singularity of $I_2$ comes entirely from the kernel $\b$ defined in
(\ref{alphabetareal}). Therefore, to prove the IR safety of
$\bar\G_n^{NG}$ it is sufficient to focus on the part of
Eq.~(\ref{GNNG}) containing $\b$.
Splitting the arguments of the vertex into $m$ hard and $n-m$ soft modes and
collecting the dangerous contributions we obtain,
\begin{multline}
\label{eq:poles}
 \bar\G'^{NG}_{n}(\k_1,...,\k_{m},\q_1,...,\q_{n-m})\\
 =\frac{-2}{n-3}\sum_{j=1}^{n-m}\left[\sum_{i=1}^{m}
 \beta(\k_i,\q_j) \right]\bar\G'^{NG}_{n-1}(\k_1,...,\k_{m},\q_1 ,..., \check{\q}_j ,...,\q_{n-m})+O(\varepsilon^0)\,.
 \end{multline}
The sum in the square brackets is 
\be 
\sum_{i=1}^{m}
 \beta(\k_i,\q_j)
=\frac{\left(\q_j\cdot \sum_{i=1}^{m}\k_i\right)}{2q_j^2}+O(\varepsilon^0)\,.
\ee
Due to momentum conservation the sum of all the arguments entering the
vertex $\bar\G'^{NG}_{n}$ is zero,  
which implies 
\be 
\sum_{i=1}^{m}\k_i=-\sum_{j=1}^{n-m}\q_j=O(\varepsilon)\quad \Longrightarrow \quad \sum_{i=1}^{m}
 \beta(\k_i,\q_j)=O(\varepsilon^0)\,.
\ee
We conclude that the poles in the second line of \eqref{eq:poles} cancel,
and the vertex $\bar\G'^{NG}_{n}$ is IR safe.

\section{IR resummation of primordial oscillatory features}
\label{sec:IRres}

A well-known effect in Gaussian cosmological perturbation theory is
the suppression of BAO due to non-linear mode coupling. Physically, it
originates from large bulk flows that tend to smear short-scale
features in the correlation functions. TSPT provides an accurate
description of this effect via a systematic resummation of IR enhanced
contributions both in real \cite{Blas:2016sfa} and redshift
\cite{Ivanov:2018gjr} spaces. In this section we show that similar
resummation technique can be developed within TSPT for the case of
oscillating features in primordial statistics. 

As a concrete example
we consider resonant power
spectrum and bispectrum 
 produced in the axion monodromy
inflation \cite{Flauger:2010ja,Cabass:2018roz},  
\bseq
\label{AMPSBS}
\begin{align}
\label{AMPS}
&P_\zeta(k) = \frac{{\cal
    A}_\zeta^2}{k^3}\left(\frac{k}{k_*}\right)^{n_s-1} 
\left[
1+ \delta n_s\cos\left(\gamma\ln \frac{k}{k_*}+\varphi_*
\right)\right]\,,\\
\label{AMBS}
& B_\zeta(k_1,k_2,k_3) =  f_{NL}^{\rm res}\frac{4{\cal
    A}_\zeta^4}{k_1^2k_2^2k_3^2}
\bigg[\sin\left(\gamma \ln \frac{k_t}{k_*}\right)
+\frac{1}{\gamma}\sum_{i\neq j}\frac{k_i}{k_j}
\cos\left(\gamma \ln \frac{k_t}{k_*}\right)
+ O\left(\frac{1}{\gamma^2}\right)\bigg]\,,
\end{align}
\eseq
where $k_t \equiv k_1+k_2+k_3$ is the sum of wavenumbers, $\varphi_*$
is a constant phase whose value is not relevant for us, and 
\be
\label{fNLdns}
f_{NL}^{\rm res}=\frac{\g^2\delta n_s}{8}\;.
\ee
The parameter $\g$ is related to the inflationary slow-roll parameter
$\epsilon_1=-\dot H_{inf}/H^2_{inf}$, the Planck mass $M_P$
and the axion decay constant
$f_a$ as 
$\gamma =\sqrt{2\epsilon_1} M_P/f_a \gg 1$; 
it is typically large, $\g\gg 1$. 
On the other hand, the second parameter of the model $\delta n_s$ is
assumed to be small. Planck constraints on oscillating features in the
power spectrum \cite{Ade:2015lrj} set an approximate bound
\cite{Cabass:2018roz} $\delta n_s\lesssim 2\cdot10^{-3}\,\g^{0.63}$ in
the range $5\lesssim \g\lesssim 10^3$. An interesting feature of the
resonant non-Gaussianity is that its amplitude can be naturally large
even when the corrections to the power spectrum are small: the value of
$f_{NL}^{\rm res}$ as high as 
$2\cdot 10^4$ is allowed 
for
$\g\sim 10^3$. Note that though the second term in (\ref{AMBS}) is
suppressed by $1/\g$, it is important to recover the correct squeezed
limit of the bispectrum \cite{Flauger:2010ja}, so we
keep it in our analysis.

Linear matter power spectrum and bispectrum are obtained by
multiplying (\ref{AMPSBS}) with the transfer function, see
Eqs.~(\ref{earlyps}), (\ref{earlybis}). This superimposes on top of
the primordial oscillations the standard BAO. Due the
smallness of the BAO component in the transfer function, its
consequences can be studied separately from the primordial
features. The BAO contribution to various cosmological
correlation functions, including damping by large bulk flows in
Gaussian perturbation theory, was analyzed using TSPT approach in
\cite{Blas:2016sfa,Ivanov:2018gjr}. 
In principle, in the non-Gaussian case BAO will be imprinted in all
linear statistics. For example, the linear bispectrum will contain BAO
wiggles even if the primordial bispectrum is smooth. Similarly to the
Gaussian case, the amplitude of these wiggles will be damped by large
bulk flows. For
completeness, we study this effect in Appendix~\ref{app:bao}. In the main text
we focus on oscillations that are already present in the primordial
statistics. We work in real space, leaving the generalization to
redshift space for future.

\subsection{Power spectrum}
\label{sec:6ps}

To warm up, we first neglect the bispectrum and study the effect of IR
modes on the primordial oscillations in the power spectrum.
The analysis closely parallels the BAO resummation in
Ref.~\cite{Blas:2016sfa}. We decompose the linear power spectrum into
oscillatory (`wiggly') and smooth (`non-wiggly') parts,
\be
\label{PSdecomp}
P_L(k)=P_L^{\rm nw}(k)+P_L^{\rm w}(k)\;,
\ee 
where, according to (\ref{AMPS}),
\be
\label{Pwiggly}
P_L^{\rm w}(k)=P_L^{\rm nw}(k)\cdot\delta n_s
\cos\left(\gamma\ln \frac{k}{k_*}+\varphi_*\right).
\ee
The Gaussian vertices $\bar\G_n$ are functionals of the
linear power spectrum. By expanding them to the first order in
$P_L^{\rm w}$ we separate them into non-wiggly and wiggly parts,
\be
\label{Gndecomp}
\bar\G_n(\narg)=\bar\G_n^{\rm nw}(\narg)+\bar \G_n^{\rm w}(\narg)\;.
\ee 
As in Sec.~\ref{sec:ir}, let us split the momenta into hard $\{\k_i\}$ and
soft $\{\q_j\}$ and analyze the structure of the wiggly vertices in
the limit $\ve\sim q/k\ll 1$. Consider first the cubic vertex. The
leading contribution in the
soft limit is
\be 
\label{eq:gaussianG3}
\bar\Gamma'^{\rm w}_3(\k,\q,-\k-\q)\simeq
\frac{(\q\cdot \k)}{q^2}\,\frac{P_{L}^{\rm w}(|\k+\q|)
-P_{L}^{\rm w}(k)}{\big(P_L^{\rm nw}(k)\big)^2}\,.
\ee
Consistently with the results of Sec.~\ref{sec:ir}, the pole of the
first factor cancels with the numerator of the second factor at $q\to
0$. However, the cancellation does not occur if $q$ is bigger than the
period of oscillations, $q\gtrsim k/\g$, in which regime the
contribution is enhanced by $1/\ve$. To keep the track of this
enhancement, we introduce the linear operator $\D_\q$ acting on the
wiggly power spectrum, 
\be
\label{eq:defD}
\D _\q P_L^{\rm w}(\k)= 
\frac{(\q\cdot \k)}{q^2}\big(P_{L}^{\rm w}(|\k+\q|)-P_L^{\rm
    w}(k)\big)
=\frac{(\q\cdot \k)}{q^2}\big(e^{\q\cdot \nabla_{\k'}}-1\big)P_{L}^{\rm w}(k')
\Big|_{\k'=\k}\,.
\ee
Each insertion of this operator will be treated as a quantity of order
$1/\ve$. 
Note that this operator is exactly the same as in the case of BAO
resummation \cite{Blas:2016sfa}. Next, one can show that the leading
contribution in an $n$-point vertex with $n-2$ soft momenta is of
order $(1/\ve)^{n-2}$ and has the form,
\be
\label{eq:NGosc}
\begin{split}
\bar\Gamma'^{\rm w}_{n}(\k,-\k-\q_{1\ldots(n-2)},\q_1,...,\q_{n-2})
\simeq \frac{(-1)^{n-1}}{\big(P_L^{\rm nw}(k)\big)^2}\prod_{j=1}^{n-2}\D_{\q_j}
P_L^{\rm w}(k)\,.
\end{split}
\ee
The proof of this formula is given in Appendix~\ref{app:integral}. 

We are now ready to identify the most IR enhanced diagrams among the 
loop corrections to the wiggly power spectrum. These
diagrams should contain vertices with the maximal number of legs
carrying soft momenta. 
They correspond
to \textit{daisy} graphs, obtained by dressing the wiggly vertices
with soft loops (`\textit{petals}'). 
Thus, in the leading IR order we have,
\be
\label{PSresdiag}
P^{\rm w,\,LO}_{\delta\delta}(k)= 
	\begin{fmffile}{wigglyps}
	    \begin{gathered}
        \begin{fmfgraph*}(60,60)
        \fmfpen{thick}
       \fmfleft{l1}
        \fmfright{r1}
		\fmf{wiggly,label=$ $}{l1,r1}
	    \end{fmfgraph*}
	     \end{gathered}
	    \end{fmffile} 
~+~
 	\begin{fmffile}{wiggly4}
 	    \begin{gathered}
           \begin{fmfgraph*}(80,60)
         \fmfpen{thick}
         \fmfkeep{1loop}
         \fmfleft{l1,l2,l3}
         \fmfright{r1,r2,r3}
         \fmfv{d.sh=circle,d.filled=shaded,d.si=.15w,label=$ $,l.a=-90,l.d=.0w}{b1}
 		\fmf{plain,label=$$}{b1,l2}
 		\fmf{plain,label=$ $}{b1,r2}
 \fmf{phantom}{l1,u1}
 		\fmf{phantom}{u2,r1}
 		\fmf{phantom}{l3,v1}
 		\fmf{phantom}{v2,r3}
 	    \fmf{phantom,right=0.5,tension=0.01,l.side=left}{b1,u1}
 	     \fmf{phantom,right=0.5,tension=2,label=$ $}{u1,u2}
 	   \fmf{phantom,left=0.5,tension=0.01}{b1,u2}
 	   	    \fmf{plain,left=0.5,tension=0.01,l.side=right}{b1,v1}
 	     \fmf{plain,left=0.5,tension=2,label=$ $}{v1,v2}
 	   \fmf{plain,right=0.5,tension=0.01}{b1,v2}
	    \end{fmfgraph*}
  	     \end{gathered}   
   \end{fmffile}
   ~+~
 	    \begin{fmffile}{wiggly6}
      \begin{gathered}
         \begin{fmfgraph*}(80,60)
         \fmfpen{thick}
         \fmfkeep{2loop}
         \fmfleft{l1,l2,l3}
         \fmfright{r1,r2,r3}
         \fmfv{d.sh=circle,d.filled=shaded,d.si=.15w,label=$ $,l.a=155,l.d=.0w}{b1}
 		\fmf{plain,label=$ $}{b1,l2}
 		\fmf{plain,label=$ $}{b1,r2}
 		\fmf{phantom}{l1,u1}
 		\fmf{phantom}{u2,r1}
 		\fmf{phantom}{l3,v1}
 		\fmf{phantom}{v2,r3}
 	    \fmf{plain,right=0.5,tension=0.01,l.side=left}{b1,u1}
 	     \fmf{plain,right=0.5,tension=2,label=$ $}{u1,u2}
 	   \fmf{plain,left=0.5,tension=0.01}{b1,u2}
 	   	    \fmf{plain,left=0.5,tension=0.01,l.side=right}{b1,v1}
 	     \fmf{plain,left=0.5,tension=2,label=$ $}{v1,v2}
 	   \fmf{plain,right=0.5,tension=0.01}{b1,v2}  
 	    \end{fmfgraph*}
 	    	    	    \end{gathered}
 	    \end{fmffile}
~+~\ldots
\;,
\ee
where we denoted by a wavy line and shaded circles the wiggly linear 
power spectrum and the wiggly vertices respectively. The smooth lines
represent the non-wiggly propagator $g^2P_L^{\rm nw}(k)$.
The term with $\ell$ loops in this expression is of the order 
$g^2 (g^2/\ve^2)^{\ell}$. We see that the loop
suppression represented by $g^2$ is partially compensated by the IR
enhancement, so these contributions must be resummed. 
The leading part of the diagram with $\ell$ loops reads,
\be
\begin{split}
P^{{\rm w,\,LO},\,\ell{\rm-loop}}_{\delta\delta}(k)
=&-\frac{1}{(2\ell+2)!}\cdot (2\ell+2)\cdot (2\ell+1)\cdot (2\ell-1)!!
\cdot g^2\big(P_L^{\rm nw}(k)\big)^2\\
&\times\prod_{i=1}^{\ell}\left[\int[dq_i]\,
g^2P_L^{\rm nw}(q_i)\right]
\bar\G'^{\rm w}_{2\ell+2}(\k,-\k,\q_1,-\q_1,...,\q_\ell,-\q_\ell)\\
\simeq&\;\frac{g^{2\ell+2}}{2^\ell \ell!}\;
\prod_{i=1}^{\ell}\left[\int_{q_i\leq k_S}[dq_i]\,
g^2P_L^{\rm nw}(q_i)\D_{\q_i}\D_{-\q_i}\right] P_L^{\rm w}(k)
\,,
\end{split}
\label{PSdress}
\ee  
where in passing to the second line we used the formula
(\ref{eq:NGosc}).
We have also restricted the loop integrals to the IR domain $q\leq
k_S$. The separation scale $k_S$ defining this domain must belong to
the range $k/\g<k_S<k$. Otherwise its choice is arbitrary and
represents an intrinsic freedom in the resummation scheme. We take
$k_S$ to be a constant multiple of the momentum $k$,
\be
k_S=\varkappa\cdot k\;,
\ee
and vary the proportionality coefficient
$\varkappa$ between 0.3 and 0.7. By choosing $\varkappa$ rather close to 1
we expect to take into account all relevant IR modes. The sensitivity
of the final result to the precise value of $\varkappa$ provides
an estimate of the theoretical uncertainty which is due to the fact that
one considers only the IR part of the loops and drops the
integrals over the hard momenta $q>k_S$. This sensitivity will
decrease when one includes in the calculation higher orders in the
hard loops (see below).

\begin{figure}[t]
\begin{center}
 \includegraphics[width=0.45\textwidth]{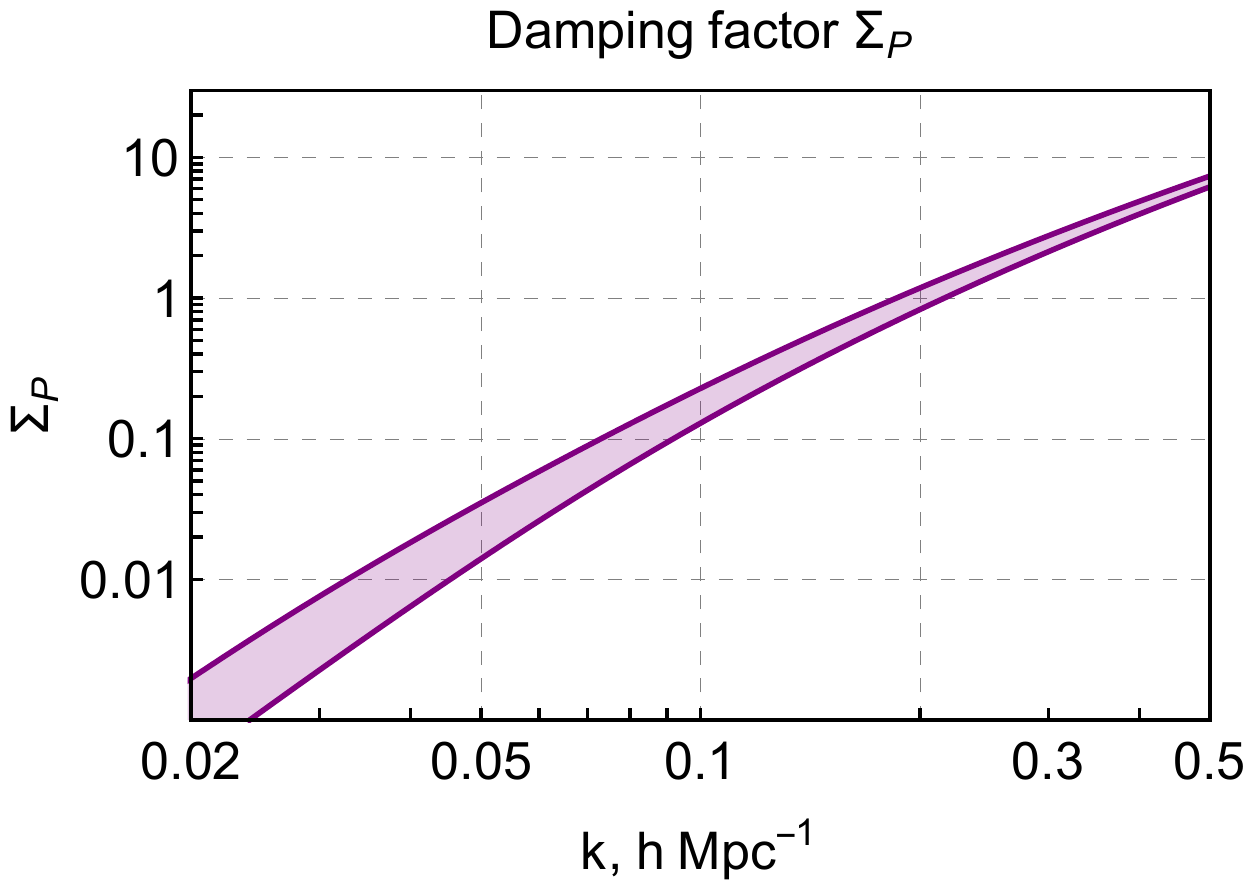}\qquad
 \includegraphics[width=0.49\textwidth]{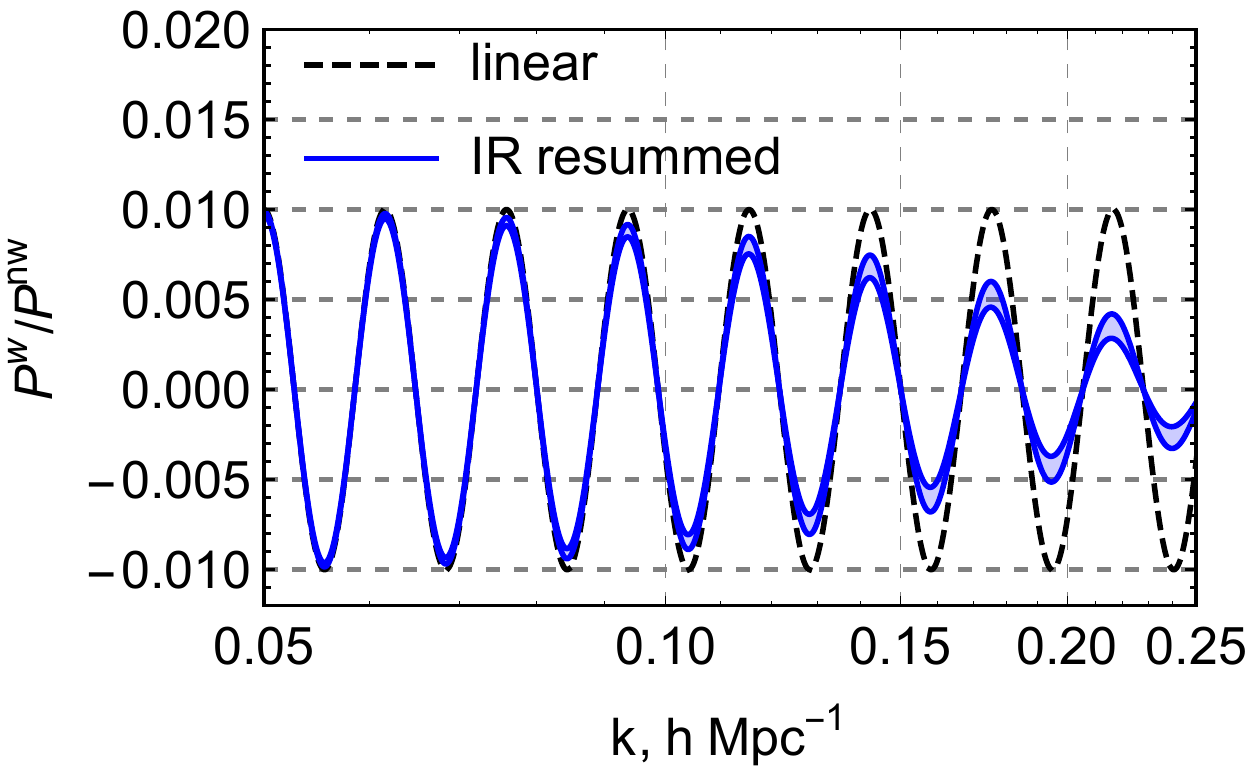}
\caption{Damping of the primordial oscillations in the power spectrum
  by large bulk flows at redshift $z=0$. {\it Left:} the damping exponent. 
{\it Right:} Comparison of the wiggly power spectrum
  before and after IR resummation. We use the value $\gamma=30$.
The bands show the change of the result when the separation scale
$k_S$ is varied in the range 
  $(0.3\div 0.7)k$.  
  \label{fig:psdamping}
}
  \end{center}
\end{figure}

Adding up the contributions (\ref{PSdress}) with all $\ell$ we obtain
the leading-order IR-resummed wiggly power spectrum in a closed form,
\be
\label{PSresummed}
P^{{\rm w,\,LO}}_{\delta\delta}(\eta;k)=g^2(\eta)
e^{-g^2(\eta){\cal S}}P^{\rm w}_L(k)\,,
\ee
where we have introduced a new operator,
\be 
\label{eq:defS}
{\cal S}=-\frac{1}{2}\int_{q\leq k_S}[dq] P_L^{\rm nw}(q)\D_\q\D_{-\q}\,.
\ee
The evaluation of its action at leading order in $\g$ and $\ve$ gives
(see Appendix~\ref{app:integral} for details),
\begin{align}
&{\cal S}P^{\rm w}_L(k)
=\Sigma_P(k;k_S)P^{\rm w}_L(k)\;,\notag\\
&\Sigma_P(k;k_S)=\frac{k^2}{6\pi^2}\int_{0}^{k_S}\!\!dq\, P_L^{\rm nw}(q)
\left[1-j_0\left(\frac{\g q}{k}\right)+2
j_2\left(\frac{\g q}{k}\right)
\right],
\label{PSdamping}
\end{align}
where $j_0$, $j_2$ are the spherical Bessel functions. Substituting
this into (\ref{PSresummed}) we see that the wiggly power spectrum is
damped. The situation is very similar to the case of BAO
\cite{Blas:2016sfa,Baldauf:2015xfa}, with the only difference that the
dependence of the damping factor $\Sigma_P$ on $k$ is now more
complicated than quadratic. We show this dependence  
in the left panel of Fig.~\ref{fig:psdamping}; the
right panel illustrates the suppression of the wiggly power spectrum.  

The total IR-resummed power spectrum at the leading order is obtained
by adding the non-wiggly contribution that remains unmodified,
\be
\label{resummedPS}
P_{\delta\delta}^{\rm LO}(\eta;k)=
g^2(\eta)\Big(P_L^{\rm nw}(k)+e^{-g^2(\eta)\Sigma_P(k)}P_L^{\rm w}(k)\Big)\;.
\ee
The resummation procedure can be extended to include the corrections
from the hard loops. We do not repeat the derivation here referring
the interested reader to \cite{Blas:2016sfa}. At the next-to-leading
order in the hard loops the result reads,
\be
\label{resummedPSNLO}
P_{\delta\delta}^{\rm NLO}=
g^2\Big(P_L^{\rm nw}
+\big(1+g^2\Sigma_P\big)e^{-g^2\Sigma_P}P_L^{\rm w}\Big)
+P_{\delta\delta}^\text{1-loop}
\Big[P_L^{\rm nw}+e^{-g^2\Sigma_P}P_L^{\rm w}\Big]
\;,
\ee
where the last term is the usual one-loop correction to the power
spectrum evaluated using the linear spectrum with damped
oscillations. Note that an extra contribution $g^2e^{-g^2\Sigma_P}P_L^{\rm w}$
in the first brackets is precisely what is needed
to avoid double counting the soft
part of the one-loop daisy diagram.

\subsection{Bispectrum}
\label{sec:6bis}

We now include into consideration primordial non-Gaussianity. We will
assume that it is purely oscillatory, as in our working example of the
axion monodromy inflation (\ref{AMBS}). In a more general case the
derivation below would apply to the wiggly part of the bispectrum. 
For clarity, we will omit the oscillations in the  
primordial power spectrum (\ref{AMPS}), 
whose effect was studied in the previous subsection,
and focus on the oscillations in the bispectrum only.


\subsubsection{Generic momenta}

We first consider the case when all three momenta in the bispectrum
are of the same order; the squeezed case with one soft momentum
will be studied afterwards. 
We start by writing the IR enhanced terms in the quartic
non-Gaussian vertex with three hard and one soft momentum. Using the
relations (\ref{eq:g3ng}) and the form of the $I_2$ kernel
(\ref{identificED2}) we obtain,
\be
\label{4NGsoft}
\begin{split}
&\bar\G_4'^{NG}(\k_1,\k_2,-\k_{12}-\q,\q)\simeq
\frac{1}{P_L(k_1)P_L(k_2)P_L(k_{12})}\\
&\quad\times\bigg[\frac{(\q\cdot\k_1)}{q^2}
\big(B_L(\k_1+\q,\k_2)-B_L(\k_1,\k_2)\big)
+\frac{(\q\cdot\k_2)}{q^2}
\big(B_L(\k_1,\k_2+\q)-B_L(\k_1,\k_2)\big)\bigg],
\end{split}
\ee 
where we temporarily adopted the representation of the linear
bispectrum as a function of two wavevectors (more precisely, of their
lengths and the relative angle). Though not symmetric with respect to
permutations in the three-point correlator, this representation turns out to be
convenient for the derivation of the resummation formula. Equation
(\ref{4NGsoft}) suggests to extend the
action of operator $\D_\q$ introduced in Sec.~\ref{sec:6ps} to the
linear bispectrum,
\be
\label{Dbis}
\D_\q B_L(\k_1,\k_2)=
\bigg[\frac{(\q\cdot\k_1)}{q^2}\big(e^{\q\cdot\nabla_{\k'_1}}-1\big)
+\frac{(\q\cdot\k_2)}{q^2}\big(e^{\q\cdot\nabla_{\k'_2}}-1\big)\bigg]
B_L(\k_1',\k_2')\Big|_{\k_1'=\k_1,\k_2'=\k_2}.
\ee
Clearly, this operator is of order $1/\ve$ in the IR power
counting. By induction one can prove an analog of Eq.~(\ref{eq:NGosc}) for the
leading IR-enhanced part of the non-Gaussian vertices (see
Appendix~\ref{app:integral}), 
\be
\label{eq:NGosc1}
\bar\Gamma'^{NG}_{n}(\k_1,\k_2,-\k_{12}-\q_{1\ldots(n-3)},\q_1,...,\q_{n-3})
\simeq
\frac{(-1)^{n-2}}{P_L(k_1)P_L(k_2)P_L(k_{12})}\prod_{j=1}^{n-3}\D_{\q_j}
B_L(\k_1,\k_2)\,.
\ee
With this result at hand, it is straightforward to identify and resum
the leading IR corrections to the non-Gaussian part of the
bispectrum. Repeating the reasoning of Sec.~\ref{sec:6ps} we obtain,
\be
\label{BISresdiag}
B^{NG,\,{\rm LO}}_{\delta\delta\delta}(\eta;\k_1,\k_2)= 
	\begin{fmffile}{bstree}
	    \begin{gathered}
        \begin{fmfgraph*}(60,60)
        \fmfpen{thick}
       \fmfleft{l1,l2,l3}
        \fmfright{r1,r2,r3}
		\fmf{plain,label=$ $}{l2,v2}
\fmf{plain,label=$ $}{v2,r2}
	\fmf{phantom,label=$ $}{l1,v1}
\fmf{phantom,label=$ $}{v1,r1}
\fmf{plain,tension=0,label=$ $}{v2,v1}
\fmfv{decor.shape=pentagram,decor.filled=empty,decor.size=7thick,label=$ $,l.a=120,l.d=.07w}{v2}
	    \end{fmfgraph*}
	     \end{gathered}
	    \end{fmffile} 
~+~
 	\begin{fmffile}{softbs5}
 	    \begin{gathered}
           \begin{fmfgraph*}(60,60)
         \fmfpen{thick}
         \fmfkeep{1loop}
         \fmfleft{l1,l2,l3}
         \fmfright{r1,r2,r3}
        \fmfv{decor.shape=pentagram,decor.filled=empty,decor.size=7thick,label=$ $,l.a=120,l.d=.07w}{b1}
 		\fmf{plain,label=$$}{b1,l2}
 		\fmf{plain,label=$ $}{b1,r2}
 \fmf{phantom}{l1,u1}
 		\fmf{phantom}{u2,r1}
 		\fmf{phantom}{l3,v1}
 		\fmf{phantom}{v2,r3}
 	    \fmf{phantom,right=0.5,tension=0.01,l.side=left}{b1,u1}
 	     \fmf{phantom,right=0.5,tension=2,label=$ $}{u1,u2}
 	   \fmf{phantom,left=0.5,tension=0.01}{b1,u2}
 	   	    \fmf{plain,left=0.5,tension=0.01,l.side=right}{b1,v1}
 	     \fmf{plain,left=0.5,tension=2,label=$ $}{v1,v2}
 	   \fmf{plain,right=0.5,tension=0.01}{b1,v2}
	\fmf{phantom,label=$ $}{l1,v3}
\fmf{phantom,label=$ $}{v3,r1}
\fmf{plain,tension=0,label=$ $}{v3,b1}
	    \end{fmfgraph*}
  	     \end{gathered}   
   \end{fmffile}
   ~+~
 	    \begin{fmffile}{softbs7}
      \begin{gathered}
         \begin{fmfgraph*}(60,60)
         \fmfpen{thick}
         \fmfleft{l0,l1,l12,l2,l223,l23,l3}
         \fmfright{r0,r1,r12,r2,l223,r23,r3}
         \fmfv{decor.shape=pentagram,decor.filled=empty,decor.size=7thick,label=$ $,l.a=120,l.d=.07w}{b1}
 		\fmf{plain,label=$ $}{b1,l2}
 		\fmf{plain,label=$ $}{b1,r2}
 		\fmf{phantom,tension=0.04}{l1,u1}
 		\fmf{phantom,tension=1}{u2,r1}
\fmf{phantom,tension=1}{w2,r12}
 		\fmf{phantom}{l3,v1}
 		\fmf{phantom}{v2,r3}
 	    \fmf{plain,right=0.3,tension=0,l.side=left}{b1,u1}
 	     \fmf{plain,right=0.35,tension=0.1,label=$ $}{u1,u2}
 \fmf{plain,right=0.35,tension=0.1,label=$ $}{u2,w2}
 	   \fmf{plain,left=0.3,tension=0.1}{b1,w2}
 	   	    \fmf{plain,left=0.5,tension=0.01,l.side=right}{b1,v1}
 	     \fmf{plain,left=0.5,tension=2,label=$ $}{v1,v2}
 	   \fmf{plain,right=0.5,tension=0.01}{b1,v2}  
	\fmf{phantom,label=$ $}{l0,v0}
\fmf{phantom,label=$ $}{v0,r0}
\fmf{plain,tension=0,label=$ $}{v0,b1}
 	    \end{fmfgraph*}
 	    	    	    \end{gathered}
 	    \end{fmffile}
~+~\ldots\;.
\ee
Evaluating the diagrams and summing them up we arrive at,
\be 
\label{BLOoper}
B^{NG,\,{\rm LO}}_{\delta\delta\delta}(\eta;\k_1,\k_2)
=g^3(\eta)e^{-g^2(\eta){\cal S}} B_L(\k_1,\k_2)\;,
\ee
where the operator ${\cal S}$ is still given by Eq.~(\ref{eq:defS}), 
with $\D_\q$ acting on the bispectrum as in
(\ref{Dbis}). Note that until now we have not used any specific
expression of the linear bispectrum, so the result of IR resummation in the
operator form (\ref{BLOoper}) is valid for any oscillating
non-Gaussianity.  

\begin{figure}
\begin{center}
    \includegraphics[width=0.55\textwidth]{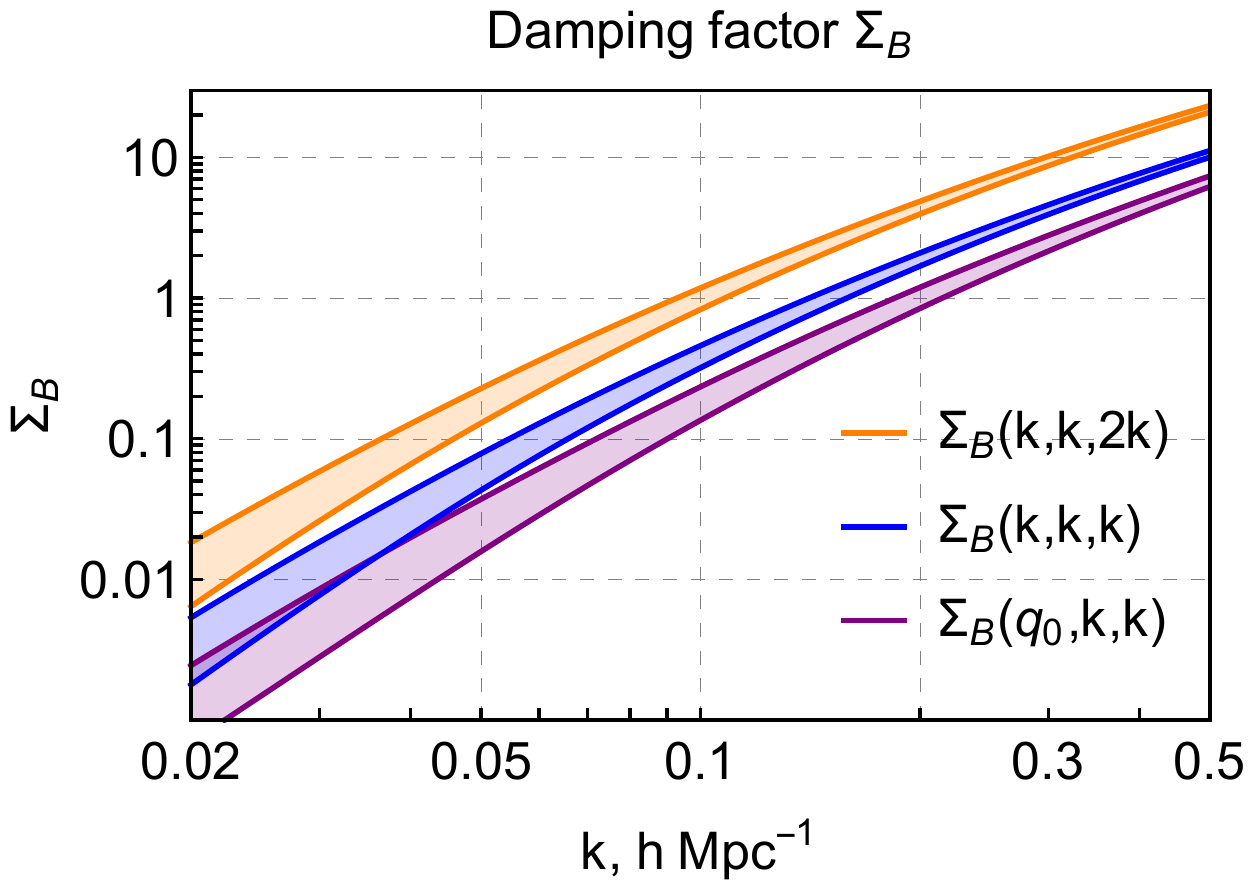}
\caption{\label{fig:damping}
The damping factor of the resonant bispectrum for flattened,
equilateral and
squeezed shapes at redshift $z=0$. We use the value $\g=30$.
The soft momentum in the squeezed case is 
$q_0=0.005\,h/\text{Mpc}$. The bands show the sensitivity of the
result to the choice of the separation scale in the range $0.15k_t\leq
k_S\leq 0.35 k_t$.
}
  \end{center}
\end{figure}

\begin{figure}[t]
\begin{center}
    \includegraphics[width=0.45\textwidth]{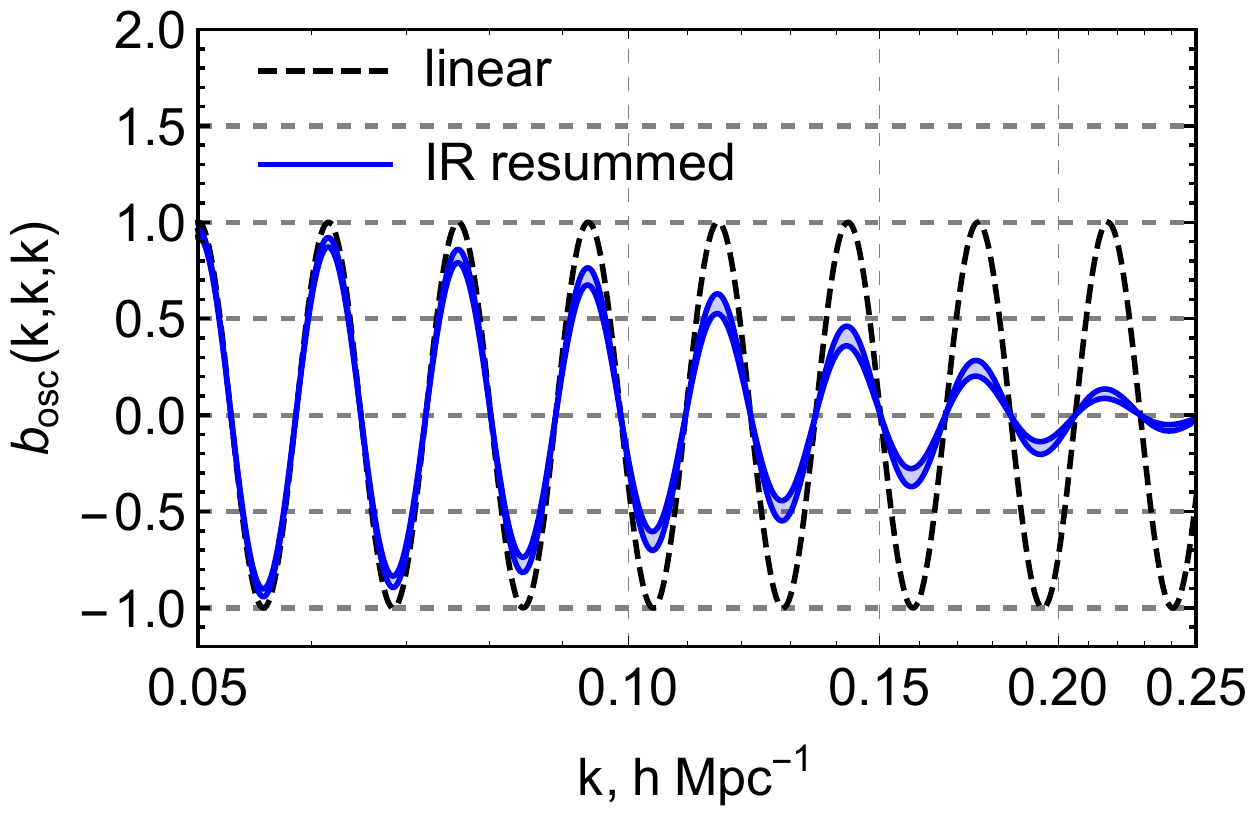}\qquad
    \includegraphics[width=0.45\textwidth]{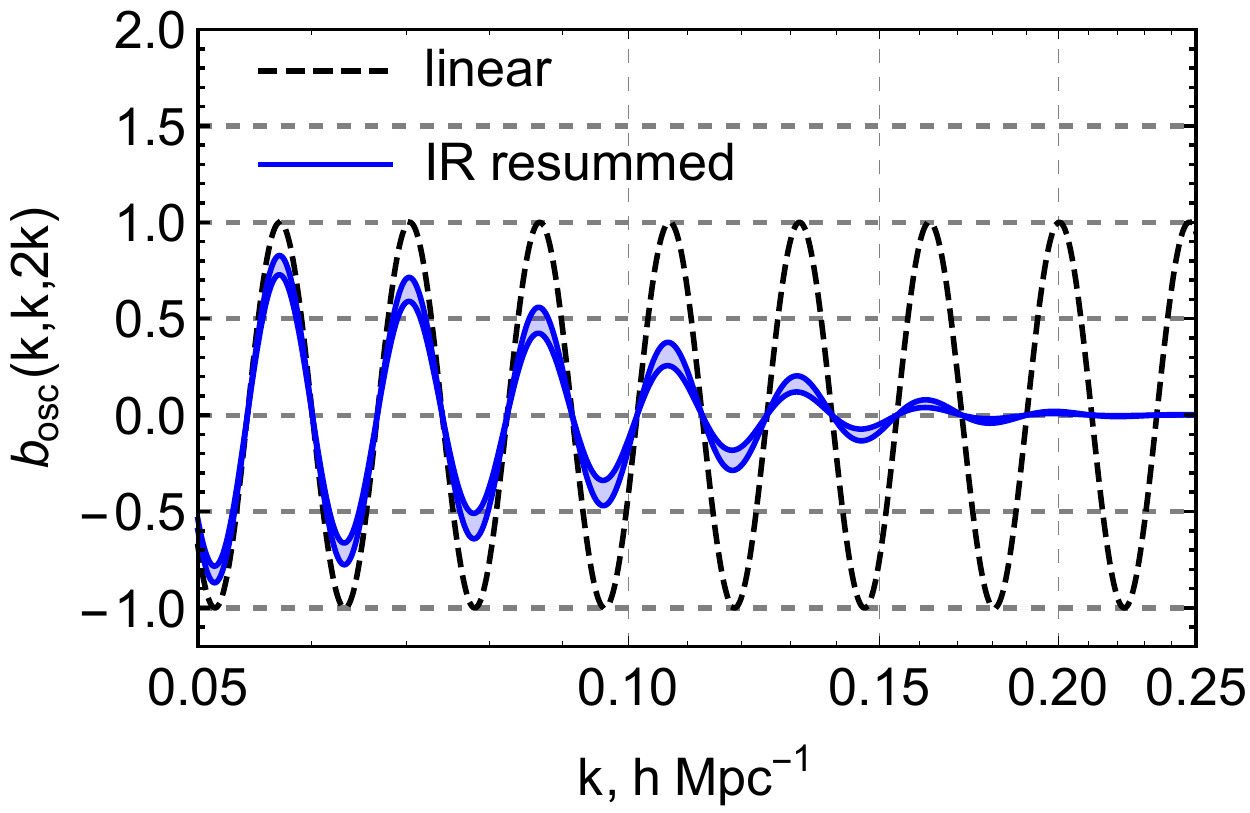}
\caption{\label{fig:plotbosc}
The oscillatory bispectrum before and after IR resummation for
equilateral ({\it left}) and flattened ({\it right}) shapes. We take $\g=30$,
$k_*=0.05\,h/\text{Mpc}$. The bands correspond to the variation of
$k_S$ in the range $(0.15\div 0.35) k_t$.
The 
redshift $z=0$.
}
  \end{center}
\end{figure}

Evaluation of the action of ${\cal S}$ leads to the damping of the
oscillations in the bispectrum,
\be 
\label{BLOexpl}
B^{NG,\,{\rm LO}}_{\delta\delta\delta}(\eta;k_1,k_2,k_3)
=g^3(\eta)e^{-g^2(\eta)\Sigma_B(k_1,k_2,k_3)} B_L(k_1,k_2,k_3)\;,
\ee 
where we have switched back to the symmetric representation of $B$ as
the function of three wavenumbers.
The concrete expression for the damping
factor depends on the model.
In
Appendix~\ref{app:integral} we perform the calculation for the case
of the resonant bispectrum (\ref{AMBS}) with the result,
\be
\label{BSdamping}
\begin{split} 
&\Sigma_B(k_1,k_2,k_3)= \sum_{i<j}^3 k_i k_j 
J\big(\hat{\bf k}_i\cdot \hat{\bf k}_j\big)\;,\\
& J (x)=
\int_{q\leq k_S} \frac{dq}{6\pi^2}P_L(q)
\left[-x+x j_0\left(\frac{\g q}{k_t}\sqrt{2(1-x)}\right)
+\frac{3-x}{2}j_2\left(\frac{\g q}{k_t}\sqrt{2(1-x)}\right)
\right]\,.
\end{split}
\ee
Here $\hat\k_i$ is a unit vector directed along the momentum $\k_i$.
One observes that, apart from the dependence on the absolute values of the momenta, the damping 
has a non-trivial
dependence on the shape of the momentum triangle. This is
illustrated in Fig.~\ref{fig:damping} where we show the dependence of
$\Sigma_B$ on the momentum for several shapes. The IR separation scale
is varied in the range $0.15k_t\leq k_S\leq 0.35k_t$.
Note that with this choice $\Sigma_B$ reduces to the power spectrum
damping $\Sigma_P(k_1)$ in the squeezed limit $k_3\ll k_1\sim k_2$
(more on the squeezed limit below).
We see that the
dependence on $k_S$ is rather mild. As noted before, it provides
an estimate of the uncertainty due to the neglect of the higher loop
corrections. 
To illustrate the effect of damping, we split the bispectrum into the
envelope\footnote{In the expression for the envelope we include the
  factor coming from the transfer functions, see
  Eq.~(\ref{earlybis}).} 
$B_{\rm env}$ and oscillatory part $b_{\rm osc}$, 
\be
B_L=B_{\rm env}\cdot b_{\rm osc}\;,
\qquad
B_{\rm env}(k_1,k_2,k_3)=4f_{NL}^{\rm res}{\cal A}_\zeta
\,\sqrt{\frac{P_L(k_1)P_L(k_2)P_L(k_3)}{k_1k_2k_3}}\;.
\label{envelop}
\ee
In Fig.~\ref{fig:plotbosc} we show the
effect of IR resummation on $b_{\rm osc}$ for the value of the
frequency parameter $\g=30$. We observe that the oscillations are
significantly damped at $k\gtrsim 0.07\,h/\text{Mpc}$ and get
essentially washed out at $k\gtrsim 0.15\,h/\text{Mpc}$.

\subsubsection{The squeezed limit}

So far we have been assuming that all three arguments of the
bispectrum are of the same order, $k_1\sim k_2\sim k_3$. This
assumption does not cover the squeezed limit when one of the momenta
is soft. The latter limit is relevant for the scale-dependent halo
bias. Besides, in the cosmological collider models of
\cite{Arkani-Hamed:2015bza,Lee:2016vti,Arkani-Hamed:2018kmz} this is precisely
the limit that encodes the information about the masses and spins 
of heavy
particles. Thus, it deserves a separate attention. We are going to see
that it leads to new IR-enhanced contributions. 

Consider again the non-Gaussian four-point vertex where now only two momenta
are hard. A brief calculation gives its leading IR-enhanced part,
\be
\label{4NGsoftsqueezed}
\bar\G_4'^{NG}(\k,-\k-\q_{12},\q_1,\q_2)\simeq
\frac{1}{\big(P_L(k)\big)^2}\bigg[\frac{1}{P_L(q_2)}\D_{\q_1}B_L(\k,\q_2)
+\frac{1}{P_L(q_1)}\D_{\q_2}B_L(\k,\q_1)\bigg],
\ee  
where the operator $\D_\q$ acts only on the hard momentum in the
linear bispectrum,
\be
\label{Dbissqueezed}
\D_\q
B_L(\k,\q')=\frac{(\q\cdot\k)}{q^2}\big(e^{\q\nabla_{\k'}}-1\big)
B_L(\k',\q')\Big|_{\k'=\k}\;.
\ee 
In Appendix~\ref{app:integral} we generalize this formula to an
arbitrary $n$-point vertex,
\be
\label{eq:NGosc2}
\bar\Gamma'^{NG}_{n}(\k,-\k-\q_{1\ldots(n-2)},\q_1,...,\q_{n-2})
\simeq
\frac{(-1)^{n-2}}{\big(P_L(k)\big)^2}
\sum_{i=1}^{n-2}\frac{1}{P_L(q_i)}\bigg(\prod_{j\neq i}^{n-3}\D_{\q_j}\bigg)
B_L(\k,\q_i)\,.
\ee
We observe that the leading IR term is of order $1/\ve^{n-3}$, as in
(\ref{eq:NGosc1}), but its structure is different. Therefore, the
leading contributions to the squeezed bispectrum are still given by
the diagrams (\ref{BISresdiag}), which, however, need to be
re-evaluated. Denoting the soft external momentum by $\q_0$ we obtain
for the $\ell$-loop term in the series,   
\be
\label{BLOl-loop}
\begin{split}
&B_{\delta\delta\delta}^{NG,\,{\rm LO},\,\ell\text{-loop}}(\k,\q_0)
\simeq 
\;\frac{g^3}{\ell!}\bigg(\frac{g^2}{2}\bigg)^\ell
\bigg(\int_{q\leq k_S}[dq]P_L(q)\D_\q\D_{-\q}\bigg)^\ell B_L(\k,\q_0)\\
&\qquad\qquad\quad
+\frac{g^5P_L(q_0)}{(\ell-1)!}\bigg(\frac{g^2}{2}\bigg)^{\ell-1}
\!\bigg(\!\int_{q\leq k_S}\!\![dq]P_L(q)\D_\q\D_{-\q}\!\bigg)^{\ell-1}
\!\!\int_{q'\leq k_S}\!\![dq']\D_{\q_0}\D_{-\q'}B_L(\k,\q')\;.
\end{split}
\ee
Summation over $\ell$ yields 
the leading order squeezed bispectrum in the operator form,
\be
\label{BLOsqueezed}
B_{\delta\delta\delta}^{NG,\,{\rm LO}}(\eta;\k,\q_0)\!=\!
g^3(\eta)e^{-g^2(\eta){\cal S}}B_L(\k,\q_0)
+g^5(\eta)P_L(q_0)\!\!\int_{q\leq k_S}\!\!\![dq]
\D_{\q_0}\D_{-\q}e^{-g^2(\eta){\cal S}}B_L(\k,\q)\,.
\ee
The first term here is the same as Eq.~(\ref{BLOexpl}). To see this we 
evaluate the action of the operator ${\cal S}$,
\be
\label{Ssqueezed}
{\cal S}B_L(\k,\q_0)=\Sigma_P(k)B_L(\k,\q_0)\;,
\ee
where $\Sigma_P$ is the power-spectrum damping factor
(\ref{PSdamping}). As discussed before, 
the latter coincides with the squeezed limit of the
bispectrum damping $\Sigma_B$.

The second contribution in
(\ref{BLOsqueezed}) is new. 
It describes the distortion of the oscillatory pattern produced by the
long mode $\q_0$. Of course, this effect is non-vanishing only if
$q_0\gtrsim k/\g$. In real space this corresponds to the requirement
that the length of the long mode should be still shorter than the
correlation length $l_{\rm corr}\sim \g/k$ set up by the
momentum-space oscillations. Spelling out the operators in
(\ref{BLOsqueezed}) we obtain an explicit expression,
\be
\label{BLOsqueezed1}
B_{\delta\delta\delta}^{NG,\,{\rm LO}}(\eta;|\k-\tfrac{\q_0}{2}|,q_0,
|\k+\tfrac{\q_0}{2}|)=B_{(1)}(\eta;\k,\q_0)+B_{(2)}(\eta;\k,\q_0)\;,
\ee
where
\bseq
\begin{align}
\label{Bsqueezed1}
B_{(1)}=&g^3(\eta)e^{-g^2(\eta)\Sigma_P(k)}B_L(|\k-\tfrac{\q_0}{2}|,q_0,
|\k+\tfrac{\q_0}{2}|)\;,\\
B_{(2)}=&2g^5(\eta)e^{-g^2(\eta)\Sigma_P(k)}P_L(q_0)
\frac{(\k\cdot\q_0)}{q_0^2}\notag\\
&\times\int_{q\leq k_S}[dq]
\frac{(\k\cdot\q)}{q^2}
\Big(B_L(|\k+\tfrac{\q_0}{2}|,q,|\k+\tfrac{\q_0}{2}+\q|)
-B_L(|\k-\tfrac{\q_0}{2}|,q,|\k-\tfrac{\q_0}{2}+\q|)\Big).
\label{Bsqueezed2}
\end{align} 
\eseq
Note that we have shifted the hard momentum $\k\mapsto
\k-\frac{\q_0}{2}$ to somewhat simplify the subsequent formulas.

Evaluation of the resonant bispectrum in the squeezed limit is
subtle. Taken at face value, the expression (\ref{AMBS}) will be
dominated at $q\lesssim k/\g$ by the second term in the square
brackets. The latter can be written as\footnote{With the appropriate
  value of the phase $\varphi_*=\pi/2$ in (\ref{AMPS}).}
\be
\label{consist}
B_\zeta\Big|_{q\to 0,\; k_1=|\k-\tfrac{\q}{2}|}
\simeq f_{NL}^{\rm res}\frac{4{\cal A}_\zeta^4}{k^4 q^2}
\cdot\frac{2k}{\g q}\cos\Big(\g\ln\frac{2k}{k_*}\Big)
=-P_\zeta(q)P_\zeta(k)\frac{d \ln(k^3 P_\zeta(k))}{d\ln k}\;,
\ee 
which is precisely of the form dictated by the Maldacena consistency
condition \cite{Maldacena:2002vr}. It is known that such terms
drop from physical observables. In particular, they cancel out if one
transforms the bispectrum from the comoving
  coordinates used to derive (\ref{AMBS}) to 
conformal Fermi coordinates (CFC) \cite{Pajer:2013ana}. In the latter
frame the leading
contribution into squeezed primordial bispectrum of a single-field
slow-roll inflation scales as $O(q^2/k^2)$. Thus, to avoid unphysical
contributions, the term (\ref{consist}) must be subtracted. On the
other hand, the construction of the CFC frame is possible only if the
wavelength of the long mode is larger than all other relevant
scales. It breaks down when the length of the long mode becomes
comparable to the correlation length\footnote{Indeed, it can be shown
  using the results of \cite{Cabass:2016cgp} 
that the terms introduced into the
  squeezed bispectrum by the transformation from the comoving frame
  to CFC are organized as an expansion in powers of 
$(q l_{\rm corr})^2$.} 
$l_{\rm corr}\sim\g/k$.  
In other words, while we can easily obtain an unambiguous physical squeezed
limit of the bispectrum at $q\ll k/\g$ using CFC, it is no longer
possible at $k/\g\lesssim q\ll k$. A proper definition of the physical
bispectrum in the intermediate range presents an open issue and may
require taking into account the projection effects pertaining to the
actual physical quantities measured by an observer on Earth. Such
study is beyond the scope of this paper. Instead, we adopt a
simplified approach and model the expected form of the physical
bispectrum by subtracting from (\ref{AMBS}) the term (\ref{consist})
weighted with a function $w\big(\frac{\g q}{2k}\big)$ that
interpolates between
$w=1+O(x^2)$ at $x\to 0$ and $w=0$ at $x\to \infty$; in the practical
calculations we will use $w=(1+x^2)^{-1}$. This model cannot be judged
fully satisfactory. In particular, it likely introduces an order-one
error compared to the true bispectrum at $\frac{\g q}{2k}\sim
1$. Nevertheless, it is sufficient for our purposes of illustrating
the dynamical effects of large bulk flows.

\begin{figure}
\begin{center}
    \includegraphics[width=0.45\textwidth]{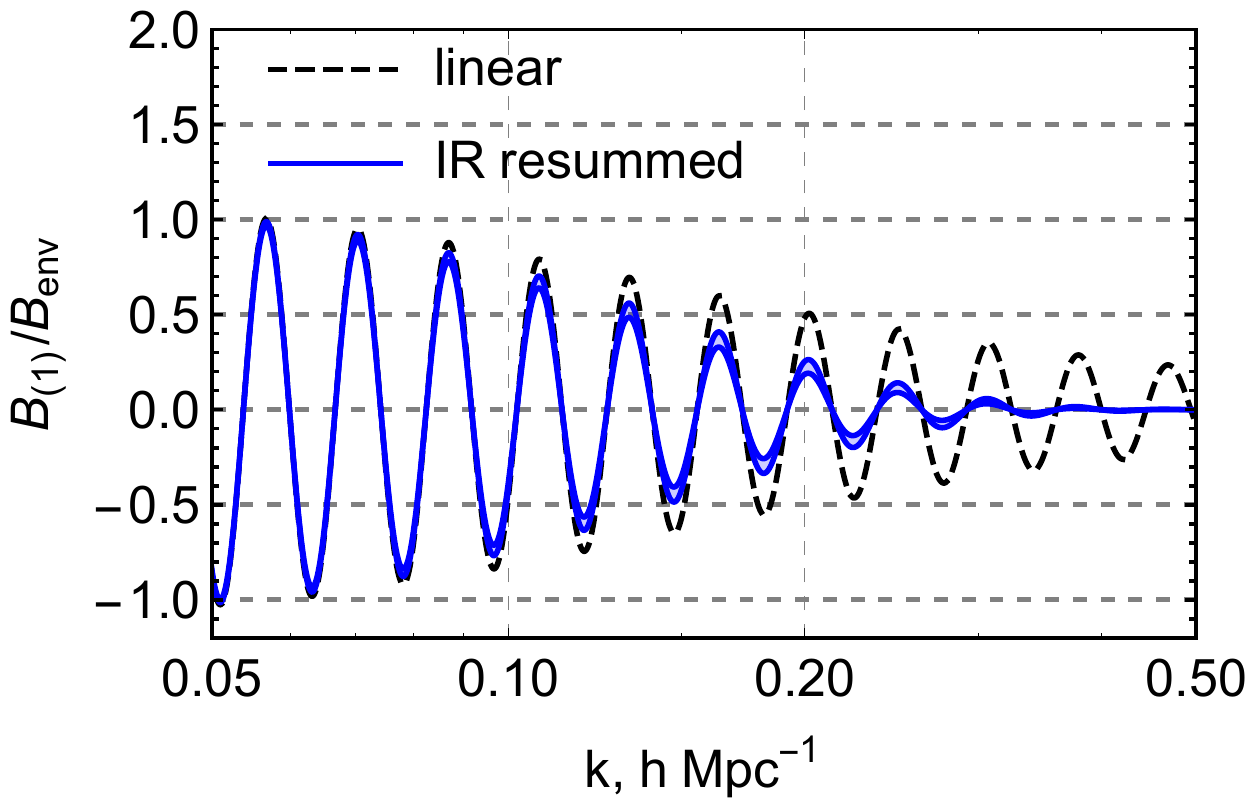}\qquad
    \includegraphics[width=0.45\textwidth]{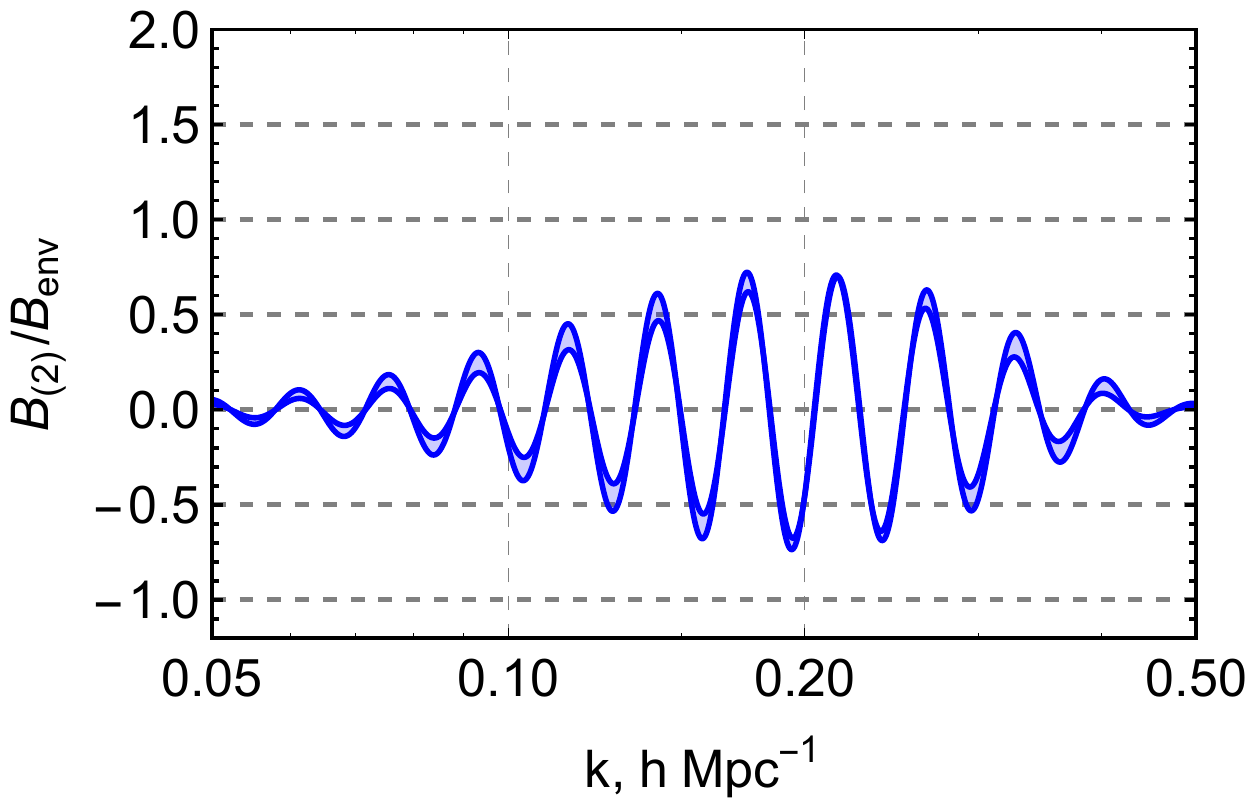}
\caption{\label{fig:squeezed}
The components $B_{(1)}$ ({\it left}) and $B_{(2)}$ ({\it right}) 
of IR resummed squeezed bispectrum as functions of the hard
momentum for the value of the soft momentum $q_0=0.005\,h/\text{Mpc}$.
The hard and soft momenta are taken to be collinear; $\g=30$, 
$k_*=0.05\,h/\text{Mpc}$ . The bands correspond to the variation of
$k_S$ in the range $(0.3\div 0.7) k$.
The 
redshift $z=0$.
}
  \end{center}
\end{figure}

\begin{figure}
\begin{center}
    \includegraphics[width=0.45\textwidth]{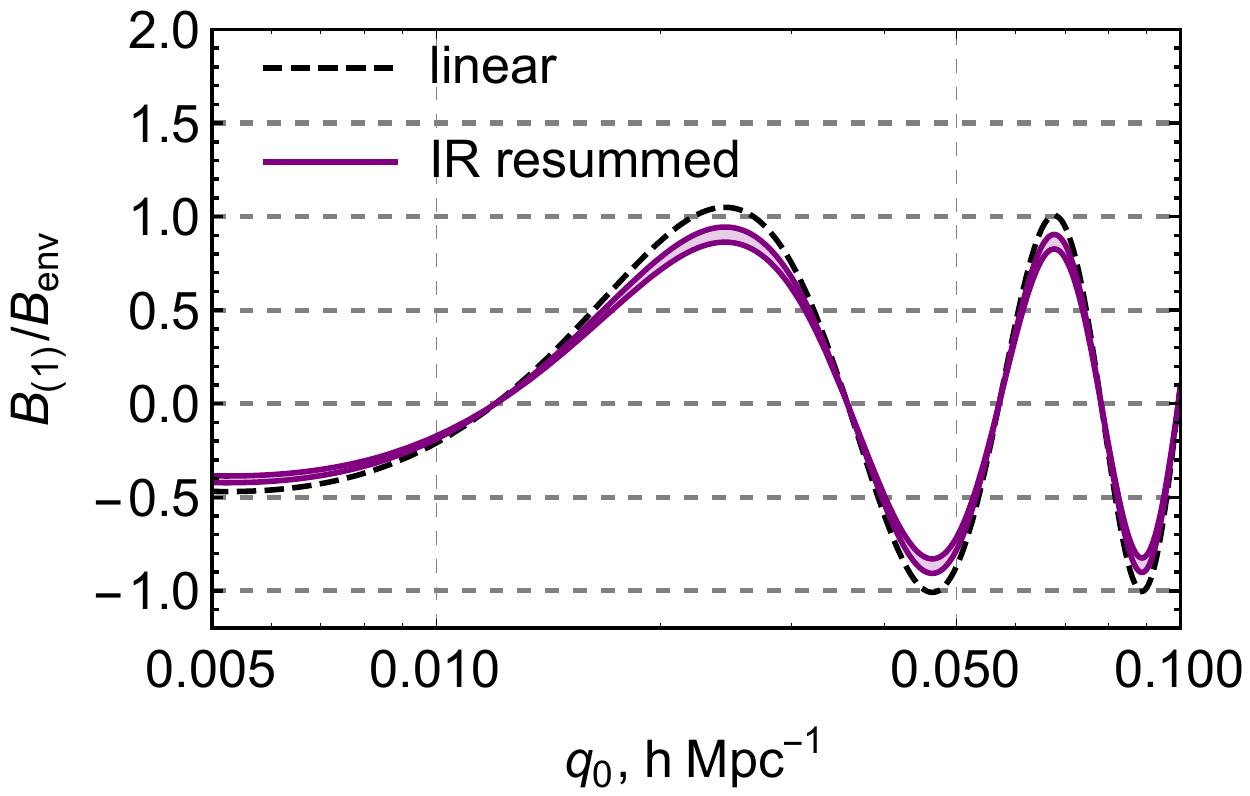}\qquad
    \includegraphics[width=0.45\textwidth]{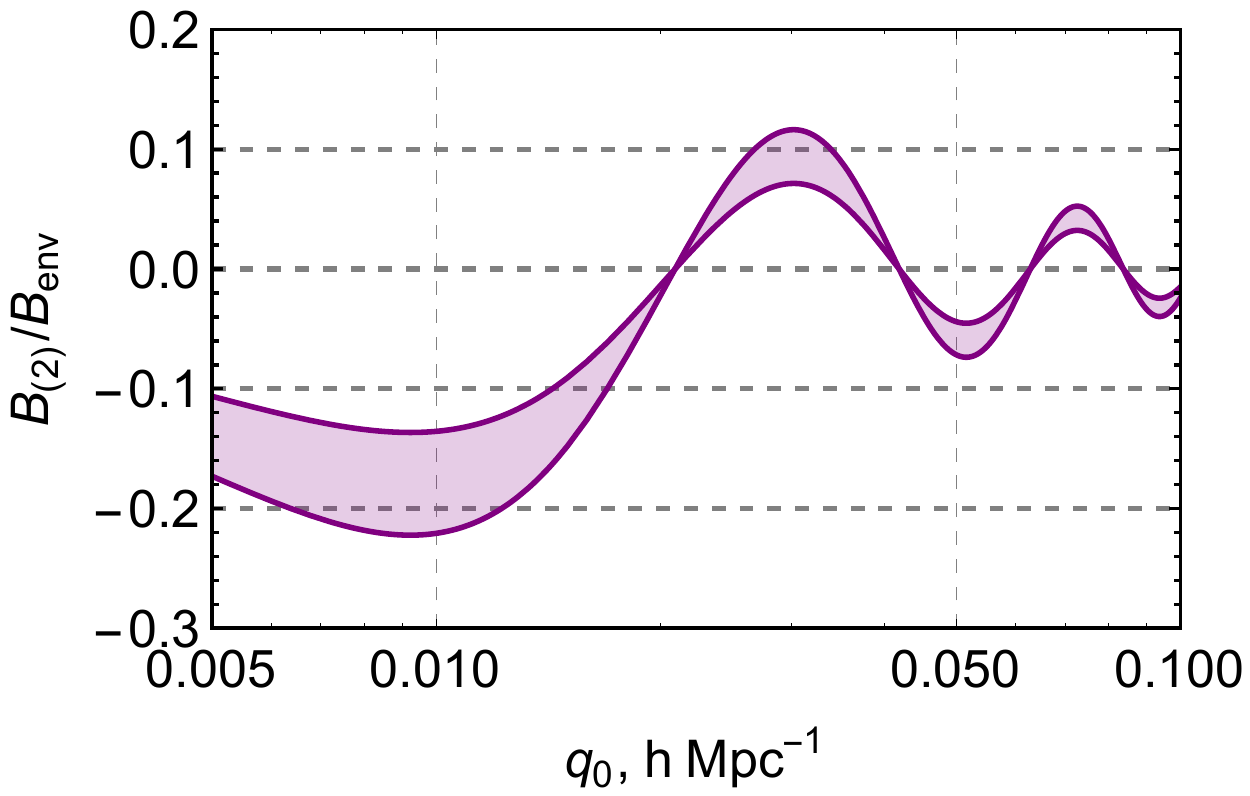}
\caption{\label{fig:squeezed1}
The components $B_{(1)}$ ({\it left}) and $B_{(2)}$ ({\it right}) 
of IR resummed squeezed bispectrum as functions of the soft
momentum for the hard momentum $k=0.1\,h/\text{Mpc}$. 
The hard and soft momenta are taken to be collinear; $\g=30$, 
$k_*=0.05\,h/\text{Mpc}$ . 
The bands correspond to the variation of
$k_S$ in the range $(0.3\div 0.7) k$.
The
redshift $z=0$.
}
  \end{center}
\end{figure}

With the above caveat in mind, we will use the following form of the
squeezed linear bispectrum,
\be
\label{Blinsqueezed}
\begin{split}
B_L(|\k-\tfrac{\q}{2}|,q,|\k+\tfrac{\q}{2}|)=&B_{\rm env}(k,k,q)
\bigg[\sin\Big(\g\ln\frac{2k}{k_*}\Big)
\bigg(\cos\Big(\frac{\g q}{2k}\Big)
-\frac{2k}{\g q}\sin\Big(\frac{\g q}{2k}\Big)\bigg)\\
&+\cos\Big(\g\ln\frac{2k}{k_*}\Big)
\bigg(\sin\Big(\frac{\g q}{2k}\Big)
+\frac{2k}{\g q}\bigg(\cos\Big(\frac{\g q}{2k}\Big)
-w\Big(\frac{\g q}{2k}\Big)\bigg)\bigg)\bigg],
\end{split}
\ee
where the smooth enveloping function $B_{\rm env}$ is defined in
(\ref{envelop}). This form directly determines the contribution
$B_{(1)}$ of the IR resummed bispectrum, which differs from the linear
bispectrum only by the damping of oscillations. To find the
second contribution $B_{(2)}$ one substitutes (\ref{Blinsqueezed})
into (\ref{Bsqueezed2}). Integration over the directions of $q$
yields,
\be
\label{B2final}
\begin{split}
\frac{B_{(2)}(\eta;\k,\q_0)}{B_{\rm env}(k,k,q_0)}=&
4g^5(\eta)e^{-g^2(\eta)\Sigma_P(k)}\,k\sqrt{q_0P_L(q_0)}\,
\frac{(\k\cdot\q_0)}{q_0^2}\sin\bigg(\g\frac{(\k\cdot\q_0)}{2k^2}\bigg)\\
&\times\bigg[C_1(k;k_S)\sin\Big(\g\ln\frac{2k}{k_*}\Big)
+C_2(k;k_S)\cos\Big(\g\ln\frac{2k}{k_*}\Big)\bigg],
\end{split}
\ee
with
\bseq
\label{C12}
\begin{align}
C_1=&\int_{q\leq k_S}\frac{dq}{2\pi^2}\sqrt{qP_L(q)}
\;j_1\Big(\frac{\g q}{2k}\Big)\bigg[\frac{2k}{\g q}
\sin\Big(\frac{\g q}{2k}\Big)-\cos\Big(\frac{\g q}{2k}\Big)\bigg],\\
C_2=&-\int_{q\leq k_S}\frac{dq}{2\pi^2}\sqrt{qP_L(q)}
\;j_1\Big(\frac{\g q}{2k}\Big)\bigg[\sin\Big(\frac{\g q}{2k}\Big)
+\frac{2k}{\g q}\bigg(\cos\Big(\frac{\g q}{2k}\Big)
-w\Big(\frac{\g q}{2k}\Big)\bigg)\bigg].
\end{align}
\eseq
Note the peculiar dependence of the expression (\ref{B2final}) 
on the
angle between the soft and hard momenta: for $q_0\gg k/\g$ it rapidly
oscillates as a function of the angle, whereas at $q_0\ll k/\g$ it
reduces to the sum of monopole and quadrupole. 
We plot the contributions $B_{(1)}$, $B_{(2)}$ (for collinear $\k$ and
$\q_0$) as functions of the hard momentum $k$ in
Fig.~\ref{fig:squeezed}. We observe the damping of oscillations in
$B_{(1)}$
above
$k\sim 0.15h/\text{Mpc}$, though the effect is somewhat weaker than
in the equilateral and flattened cases. The behavior of $B_{(2)}$ is
quite different: its amplitude grows from small $k$ up to 
$k\sim 0.2h/\text{Mpc}$ and only then starts decaying exponentially.
In Fig.~\ref{fig:squeezed1} we
plot the components of the squeezed bispectrum 
as functions of the soft momentum at fixed $k$.

Finally, let us notice that the squeezed limit of the resonant
bispectrum, through the mode coupling, produces an additional
oscillating contribution in the power spectrum on top of the
primordial one (\ref{Pwiggly}). Its leading part can be found using
again the method of IR resummation. It is given by the sum of daisy
diagrams,
\be
\Delta P^{\rm w,\,LO}_{\delta\delta}(k)= 
 	\begin{fmffile}{wiggly4bs}
 	    \begin{gathered}
           \begin{fmfgraph*}(60,60)
         \fmfpen{thick}
         \fmfkeep{1loop}
         \fmfleft{l1,l2,l3}
         \fmfright{r1,r2,r3}
\fmfv{decor.shape=pentagram,decor.filled=empty,decor.size=7thick,label=$ $,l.a=120,l.d=.07w}{b1}
 		\fmf{plain,label=$$}{b1,l2}
 		\fmf{plain,label=$ $}{b1,r2}
 \fmf{phantom}{l1,u1}
 		\fmf{phantom}{u2,r1}
 		\fmf{phantom}{l3,v1}
 		\fmf{phantom}{v2,r3}
 	    \fmf{phantom,right=0.5,tension=0.01,l.side=left}{b1,u1}
 	     \fmf{phantom,right=0.5,tension=2,label=$ $}{u1,u2}
 	   \fmf{phantom,left=0.5,tension=0.01}{b1,u2}
 	   	    \fmf{plain,left=0.5,tension=0.01,l.side=right}{b1,v1}
 	     \fmf{plain,left=0.5,tension=2,label=$ $}{v1,v2}
 	   \fmf{plain,right=0.5,tension=0.01}{b1,v2}
	    \end{fmfgraph*}
  	     \end{gathered}   
   \end{fmffile}
   ~+~
 	    \begin{fmffile}{wiggly6bs}
      \begin{gathered}
         \begin{fmfgraph*}(60,60)
         \fmfpen{thick}
         \fmfkeep{2loop}
         \fmfleft{l1,l2,l3}
         \fmfright{r1,r2,r3}
\fmfv{decor.shape=pentagram,decor.filled=empty,decor.size=7thick,label=$ $,l.a=120,l.d=.07w}{b1}
 		\fmf{plain,label=$ $}{b1,l2}
 		\fmf{plain,label=$ $}{b1,r2}
 		\fmf{phantom}{l1,u1}
 		\fmf{phantom}{u2,r1}
 		\fmf{phantom}{l3,v1}
 		\fmf{phantom}{v2,r3}
 	    \fmf{plain,right=0.5,tension=0.01,l.side=left}{b1,u1}
 	     \fmf{plain,right=0.5,tension=2,label=$ $}{u1,u2}
 	   \fmf{plain,left=0.5,tension=0.01}{b1,u2}
 	   	    \fmf{plain,left=0.5,tension=0.01,l.side=right}{b1,v1}
 	     \fmf{plain,left=0.5,tension=2,label=$ $}{v1,v2}
 	   \fmf{plain,right=0.5,tension=0.01}{b1,v2}  
 	    \end{fmfgraph*}
 	    	    	    \end{gathered}
 	    \end{fmffile}
~+~\ldots
=-g^3e^{-g^2{\cal S}}\int_{q\leq k_S}[dq]\D_{-\q}B_L(\k,\q)
\;,
\ee 
where we made use of the expression (\ref{eq:NGosc2}) for the TSPT
vertices to compute different terms in the series. The structure of
this result is intuitively clear. Working out the action of the
operators ${\cal S}$, $\D_{-\q}$ one recognizes in it the leading IR
part of the usual one-loop correction to the power spectrum (last term
in Eq.~(\ref{eq:spt1loop})), with an extra exponential
damping. Substituting the explicit form of the squeezed linear
bispectrum (\ref{Blinsqueezed}) we obtain,
\be
\label{dPwiggly}
\Delta P^{\rm w,\,LO}_{\delta\delta}(\eta;k)=
8f_{NL}^{\rm res}{\cal A}_\zeta g^3(\eta)e^{-g^2(\eta)\Sigma_P(k)}
P_{L}(k)\bigg[C_1(k)\cos\Big(\!\g\ln\frac{2k}{k_*}\Big)
-C_2(k)\sin\Big(\!\g\ln\frac{2k}{k_*}\Big)\bigg],
\ee 
where the coefficients $C_1$, $C_2$ are given in (\ref{C12}). Given
the measured value of ${\cal A}_\zeta$, this contribution is always
small. Further, the relation (\ref{fNLdns}) implies that it is
subdominant compared to the primordial wiggly power spectrum as long
as $\g<100$. Thus, in our working example of the primordial
non-Gaussianity produced in the axion monodromy inflation this
contribution appears uninteresting. However, it may be relevant in
other models where the oscillating component in the primordial power
spectrum is absent or strongly suppressed. This situation occurs, for
example, in the context of cosmological colliders
\cite{Arkani-Hamed:2015bza,Lee:2016vti,Arkani-Hamed:2018kmz}. 

\section{Conclusion}
\label{sec:conclusions}

In this paper we worked out the TSPT framework for non-Gaussian
primordial statistics. They are included by setting appropriate
initial conditions for the TSPT statistical weight, which eventually
modifies only the vertices $\G_n$. Other building blocks of TSPT, such
as the composite operator kernels $K_n$ and counterterms $C_n$ remain
intact.  
In this way TSPT can easily accommodate any sort of non-Gaussian initial conditions 
without a significant modification of the formalism.
Treating initial non-Gaussian contributions as small deformations of
the Gaussian statistical weight 
generates a new perturbative expansion on top of the usual TSPT
perturbation series. 
We have shown that the new non-Gaussian TSPT vertices are manifestly
IR safe, implying IR safety of the non-Gaussian 
TSPT 
Feynman diagrams. 

TSPT with non-Gaussian initial conditions opens a 
new way to study the effects of primordial non-Gaussianity on
LSS clustering. 
As an application, we developed a new procedure to capture the
non-linear evolution of primordial  
oscillating features that may be imprinted 
in the power spectrum and higher statistics
during inflation.
We showed that such oscillations are suppressed by nonlinear effects
of bulk motions, just like BAO. We demonstrated how to
calculate the damping factor by resummation of IR enhanced diagrams
and computed it explicitly for the resonant power spectrum and
bispectrum from axion monodromy inflation. We point out that the
damping of primordial oscillating features due to large bulk flows has
been overlooked in many studies of future survey sensitivities. Taking
it into account can significantly affect the forecasts. 
It will also be interesting to study to what extent the 
damping can be removed using the density field reconstruction technique~\cite{Eisenstein:2006nk,Padmanabhan:2008dd}.

We have found
that in the case of squeezed bispectrum the IR resummation, apart from
damping, produces an additional contribution depending on the angle
between the hard and soft momenta. It will be interesting to study
observable signatures of this term, in particular, to what extent it
can be discriminated from the primordial angular dependence of the
squeezed bispectrum in the cosmological collider models. 

The results of this work reinforce the status of IR resummation in
TSPT as a universal systematic procedure, applicable beyond the case
of Gaussian initial conditions. They can be extended in several
directions. While in the current paper we focused on the power
spectrum and
bispectrum, the IR resummation can be straightforwardly applied to
higher primordial statistics, in particular, to the trispectrum which may
possess oscillating features due to an exchange of massive particles
\cite{Arkani-Hamed:2015bza,Lee:2016vti,Arkani-Hamed:2018kmz}.  
Next, it would be interesting to incorporate the effects of redshift space distortions in the IR-resummed
non-Gaussian statistics;
this can be done along the lines of \cite{Ivanov:2018gjr}. One would
also like to extend the non-Gaussian IR resummation beyond the leading
order to include the contributions of loop diagrams with hard momenta
and subleading IR corrections, as done for the Gaussian case in
Ref.~\cite{Blas:2016sfa}. 

Last but not least, we adopted in this paper the pressureless perfect fluid
approximation. To properly renormalize the contributions of short
scales, the non-Gaussian TSPT should be embedded in the framework of
EFT of LSS \cite{Baumann:2010tm,Carrasco:2012cv,Assassi:2015jqa}. We
leave this task for future.  

\vspace{0.2cm}
\textit{Note added.} 
When this paper was completed, we became aware of the work~\cite{Beutler:2019ojk}
that also discusses the damping of the primordial oscillation features 
in the power spectrum. When overlap, our results agree. We thank the authors 
of~\cite{Beutler:2019ojk} for sharing with us the preliminary version of the paper.

\section*{Acknowledgments}

We thank Diego Blas, Mathias Garny, Azadeh Moradinezhad Dizgah, Alvise
Raccanelli, Marko Simonovi\'c, 
Fabian Schmidt, Zvonimir Vlah and Benjamin Wallisch for useful
discussions. 
M.I. thanks the CERN Theory Department for hospitality during the
completion of this work.
M.I. acknowledges the support by the RFBR grant 17-02-01008.
The work of S.S. is supported by the Swiss National Science Foundation
and the RFBR grant 17-02-00651. The work of A.V. is supported by the 
Danish National Research Foundation (DNRF91). 
M.I. is partially supported by the Simons Foundation's \textit{Origins of the Universe} program.

\appendix 

\section{TSPT kernels}
\label{app:tsptvert}

The basic building blocks of TSPT are the kernels $I_n$ and $K_n$
defined in Eqs.~(\ref{eq:psi1}), (\ref{eq:In}). They satisfy a set of
recursion relations \cite{Blas:2015qsi} 
that are derived from the dynamical equations of
motion (\ref{eq:eomsreal}). The seeds for these relations are
$I_1=K_1=1$, as follows from the linear evolution, assuming adiabatic
initial conditions. By substituting the representation (\ref{eq:psi1})
into the Euler equation (\ref{eq:eomsreal2}) and comparing with
(\ref{eq:In}) we express the kernels $I_n$ in terms of $K_n$,
\bseq
\label{identificED}
\begin{align}
\label{identificED2}
&I_2(\k_1,\k_2)=2\b(\k_1,\k_2)+\frac{3}{2}K_2(\k_1,\k_2)\;,\\
&I_n(\k_1,\ldots,\k_n)=\frac{3}{2}K_n(\k_1,\ldots,\k_n)~,~~~~n\geq 3\;.
\end{align}
\eseq
Next, the expressions (\ref{eq:psi1}), (\ref{eq:In}) are put into
the continuity equation
(\ref{eq:eomsreal1}). Comparing the terms of the same order in
$\T(\k)$ and using the relations (\ref{identificED}) we obtain the
recursive formulas,
\bseq
\label{eq:recGED}
\begin{align}
\label{K2ED}
K_2(\k_1,\k_2)=\frac{2}{7}\Big(&\a(\k_1,\k_2)+\a(\k_2,\k_2)-2\b(\k_1,\k_2)\Big)
=\frac{4}{7}\sin^2(\k_1,\k_2)\;,\\
K_n(\k_1,...,\k_n)=&
\frac{2}{2n+3}\bigg[ \sum_{i=1}^{n}
\a\big(\k_i,\k_{1\ldots n}-k_i\big)\,
K_{n-1}\big(\k_1,...,\check{\k}_i,...,\k_{n}\big)\notag\\
&-2\sum_{i<j}\b(\k_i,\k_j)\, 
K_{n-1}\big(\k_i+\k_j,\k_1,...,\check\k_i,...,\check\k_j,...,\k_n\big)\notag\\
&-\frac{3}{2} \sum_{m=2}^{n-1}\sum_{\s}
\frac{K_{m}\big(\k_{\s(1)},...,\k_{\s(m)}\big)}{m!(n-m)!}\,K_{n-m+1}
\Big(\sum_{i=1}^m\k_{\s(i)},\k_{\s(m+1)},...,\k_{\s(n)}\Big)
\Bigg].\label{Kned}
\end{align}
\eseq
The notation $\check \k_i$ above means that the momentum $\k_i$ is
absent from the arguments of the corresponding function, and in the
last line of (\ref{Kned}) the summation is performed over all
permutations $\sigma$ of $n$ indices.

\section{One-loop non-Gaussian power spectrum}
\label{app:1loopps}

In this Appendix we illustrate the application of TSPT Feynman rules
on the example of 1-loop corrections to the power spectrum involving
primordial non-Gaussianity. The corresponding diagrams are shown in
(\ref{eq:Ddelta}). To evaluate them, we need, in addition to the cubic
non-Gaussian vertex (\ref{G3NG}), also the expressions for 
the cubic Gaussian and quartic non-Gaussian vertices. These are easily
obtained using the recursion relations (\ref{Gaussrecurs}),
(\ref{GNNG}) and read,
\begin{align}
&\bar\G_3(\k_1,\k_2,\k_3)=-(2\pi)^3\delta_D^{(3)}(k_{123})\bigg[
\frac{I_2(\k_1,\k_2)}{P_L(k_3)}+\text{cyclic}\bigg]\;,\\
&  \bar\G_{4}^{NG}(\gthreearg,\k_4)=(2\pi)^3\dirdel_D(\k_{1\ldots 4})  
  \!\!\sum_{1\leq i < j \leq 4}\!\! I_2({\k_i},{\k_j})
\frac{B_L\big(\abs{\k_i+\k_j},k_l,k_m\big)}{P_L(\abs{\k_i+\k_j})P_L(k_l)P_L(k_m)} 
\bigg|_{\begin{smallmatrix}
  & l<m \\ & l,m \neq i,j \end{smallmatrix}} \;,
  \label{eq:1loopvertices}
\end{align}
Now we are ready to write the expressions corresponding to the three
diagrams in (\ref{eq:Ddelta}). For the first diagram, which we call
`daisy', we obtain,
\be 
\begin{split}
  P_{\Theta\Theta,\,{\rm daisy}}^{NG,\,\text{1-loop}}& = 
-\frac{1}{2}g^3(\eta)\big(P_L(k)\big)^2 
\intq P_L(q) {\bar\G_4'^{NG}}(\q,-\q,\k,-\k)\\
 & = -2g^3(\eta) P_L(k) \intq B_L(k,q,|\k+\q|)\frac{I_2(\k,\q)}{P_L(|\k+\q|)}\,.
 \end{split}
\ee
The second diagram (`fish') gives,
\begin{equation}
\begin{split}
  P_{\Theta\Theta,\,{\rm fish}}^{NG,\,\text{1-loop}}&=g^3(\eta)
\big(P_L(k)\big)^2\int[dq]\bar\G_3'^{NG}(\k,\q,-\k-\q)
\bar\G_3'(\k+\q,-\q,-\k)P_L(q)P_L(|\k+\q|)\\
&=g^3(\eta) 
\intq B_L(k,q,|\k+\q|) \left[\frac{2I_2(\k,\q) P_L(k)}{
        P_L(|\k+\q|)}+I_2(\k+\q,-\q)\right]. 
\end{split}
\end{equation}
In deriving these expressions we used the following properties of the
kernel $I_2$,
\[
I_2(\k,-\k)=0~,~~~~~I_2(-\k,-\q)=I_2(\k,\q)\;.
\]
Finally, the last diagram (`composite fish') reads,
\begin{equation}
\begin{split}
 P_{\delta\delta,\,\text{comp.\,fish}}^{NG,\,\text{1-loop}} 
&=-g^3(\eta)P_L(k)\int [dq] K_2(\k+\q,-\q)
\bar\G_3'^{NG}(\k+\q,-\q,-\k) P_L(q) P_L(|\k+\q|)\\
&=g^3(\eta)\intq B_L(k,q,|\k+\q|) 
 K_2(\k+\q,-\q)\;.
\end{split}
\end{equation}
Collecting the three contributions together we arrive at
\begin{equation}
\label{PddTSPT}
  P_{\delta\delta}^{NG,\,\text{1-loop}} = g(\eta)^3\intq 
B_L(k,q,\abs{\k+\q})\big(I_2(\k+\q,-\q)+K_2(\k+\q,-\q)\big)\;.
 \end{equation} 
The kernels $I_2$, $K_2$ are to be taken from
Eqs.~(\ref{identificED2}), (\ref{K2ED}). Using the relation between
the TSPT and SPT kernels established in \cite{Blas:2015qsi},
$I_2+K_2=2F_2$,
one recognizes in (\ref{PddTSPT}) the standard SPT result, 
see Eq.~(\ref{eq:spt1loop}).

\section{Technical details of IR resummation}
\label{app:integral}

\subsection{IR enhancement of the TSPT vertices}

In this subsection we derive Eqs.~(\ref{eq:NGosc}), (\ref{eq:NGosc1}),
(\ref{eq:NGosc2}). The derivation is done by induction. Let us start
with the Gaussian vertices. Assume that Eq.~(\ref{eq:NGosc}) holds for
all vertices with $n'\leq n-1$ and consider the recursion relation
(\ref{Gaussrecurs}) for the $n$-th vertex:
\be
\label{derGauss1}
\begin{split}
\bar\Gamma'^{\rm w}_{n}(\k,-\k-&\q_{tot},\q_1,...,\q_{n-2})
=-\frac{1}{n-2}\bigg[I_2(\k,-\k-\q_{tot})
\bar\G_{n-1}'^{\rm w}(\q_{tot},\q_1,...,\q_{n-2})\\
&+\sum_{i=1}^{n-2}I_2(\k,\q_i)
\bar\Gamma'^{\rm w}_{n-1}(\k+\q_i,-\k-\q_{tot},\q_1,...,
\check{\q}_i,...,\q_{n-2})\\
&+\sum_{i=1}^{n-2}I_2(-\k-\q_{tot},\q_i)
\bar\Gamma'^{\rm w}_{n-1}(-\k-\q_{tot}+\q_i,\k,\q_1,...,
\check{\q}_i,...,\q_{n-2})\\
&+\sum_{i<j}^{n-2}I_2(\q_i,\q_j)
\bar\Gamma'^{\rm w}_{n-1}(\q_i+\q_j,\k,-\k-\q_{tot},\q_1,...,
\check{\q}_i,...,\check{\q}_j...,\q_{n-2})
+\ldots\bigg],
\end{split} 
\ee
where we have denoted by $\q_{tot}$ the sum of all soft momenta. 
Dots in the last line stand
for the terms involving the vertices $\bar\G_{n'}'^{\rm w}$ with
$n'\leq n-2$ and kernels $I_m$ with $m\geq 3$. All the latter kernels
are IR safe, hence the omitted terms are at most of order
$1/\ve^{n-4}$. At the leading IR order they can be neglected. The
kernel $I_2$ has a singularity only if one of its momenta is hard and
the other is soft. Otherwise it is of order one. This implies that the
terms in the first and last lines of (\ref{derGauss1}) are at most of
order $1/\ve^{n-3}$ and can be neglected as well. One is left with the
contributions of the second and third lines. The next step is to use
the IR asymptotics 
\be
\label{I2IR}
I_2(\k,\q)\simeq \frac{(\q\cdot\k)}{q^2}
\ee
and the expression (\ref{eq:NGosc}) for $\bar\G_{n-1}'^{\rm w}$. This
yields,
\be
\label{derGauss2}
\begin{split}
\bar\Gamma'^{\rm w}_{n}(\k,-\k-\q_{tot},\q_1,...,\q_{n-2})
\simeq
\frac{(-1)^{n-1}}{(n-2)\big(P_L^{\rm nw}(k)\big)^2}\bigg[
&\sum_{i=1}^{n-2}\frac{(\q_i\cdot\k)}{q_i^2}
\prod_{j\neq i}^{n-2}\D_{\q_j}P_L^{\rm w}(|\k+\q_i|)\\
&-\sum_{i=1}^{n-2}\frac{(\q_i\cdot\k)}{q_i^2}
\prod_{j\neq i}^{n-2}\D_{\q_j}P_L^{\rm w}(\k)\bigg].
\end{split}
\ee
The terms in the square brackets combine to produce the operator $\D_{\q_i}$. 
Using the commutativity of operators $\D_\q$ with different $\q$'s one
arrives at Eq.~(\ref{eq:NGosc}). $\blacksquare$

We now turn to the proof of Eq.~(\ref{eq:NGosc1}). We again assume
that it holds for $n'\leq n-1$ and consider the relation (\ref{GNNG})
defining the $n$-th vertex. Keeping only the terms containing
the IR-enhanced kernels $I_2$ we obtain,
\be
\label{GNG3hard}
\begin{split}
&\bar\G_n'^{NG}(\k_1,\k_2,-\k_{12}-\q_{tot},\q_1,...\q_{n-3})\\
&=
-\frac{1}{n-3}
\bigg[\sum_{i=1}^{n-3}I_2(\k_1,\q_i)
\bar\G_{n-1}'^{NG}(\k_1+\q_i,\k_2,-\k_{12}-\q_{tot},
\q_1,...\check{\q}_i,...\q_{n-3})\\
&\quad+\sum_{i=1}^{n-3}I_2(\k_2,\q_i)
\bar\G_{n-1}'^{NG}(\k_2+\q_i,\k_1,-\k_{12}-\q_{tot},
\q_1,...\check{\q}_i,...\q_{n-3})\\
&\quad+\sum_{i=1}^{n-3}I_2(-\k_{12}-\q_{tot},\q_i)
\bar\G_{n-1}'^{NG}(-\k_{12}-\q_{tot}+\q_i,\k_1,\k_2,
\q_1,...\check{\q}_i,...\q_{n-3})+\ldots\bigg],
\end{split}
\ee
where dots correspond to the terms that are at most of order
$1/\ve^{n-4}$. Using next Eq.~(\ref{I2IR}) and substituting the
leading IR part of $\bar\G_{n-1}'^{NG}$ we get,
\be
\begin{split}
&\bar\G_n'^{NG}(\k_1,\k_2,-\k_{12}-\q_{tot},\q_1,...\q_{n-3})
\simeq \frac{(-1)^{n-2}}{(n-3)P_L(k_1)P_L(k_2)P_L(k_{12})}\\
&\times\! \sum_{i=1}^{n-3}\prod_{j\neq i}^{n-3}\D_{\q_j}
\bigg[\frac{(\q_i\cdot\k_1)}{q_i^2}B_L(\k_1\!+\!\q_i,\k_2)
+\frac{(\q_i\cdot\k_2)}{q_i^2}B_L(\k_1,\k_2\!+\!\q_i)
-\frac{(\q_i\cdot\k_{12})}{q_i^2}B_L(\k_1,\k_2)\bigg].
\end{split}
\ee 
Recalling that $\k_{12}=\k_1+\k_2$ we recognize in the square brackets
the operator $\D_{\q_i}$ which completes the product of operators
to $\prod_{j=1}^{n-3}\D_{\q_j}$. Summation over $i$ then cancels the
factor $(n-3)$ in the denominator and we obtain
(\ref{eq:NGosc1}). $\blacksquare$ 

Finally, we give the proof of Eq.~(\ref{eq:NGosc2}). The strategy is
the same as in the two previous cases: we assume that it is valid for
$n'\leq n-1$ and study the vertex $\G_{n}'^{NG}$. The leading
contribution containing the IR-enhanced $I_2$ kernels is
\be
\label{GNG2hard}
\begin{split}
\bar\G_n'^{NG}(\k,&-\k-\q_{tot},\q_1,...\q_{n-2})\\
=&
-\frac{1}{n-3}
\bigg[\sum_{i=1}^{n-2}I_2(\k,\q_i)
\bar\G_{n-1}'^{NG}(\k+\q_i,-\k-\q_{tot},
\q_1,...\check{\q}_i,...\q_{n-2})\\
&+\sum_{i=1}^{n-2}I_2(-\k-\q_{tot},\q_i)
\bar\G_{n-1}'^{NG}(-\k-\q_{tot}+\q_i,\k,
\q_1,...\check{\q}_i,...\q_{n-2})+\ldots\bigg].
\end{split}
\ee
Substituting the IR asymptotics of $I_2$ and $\bar\G_{n-1}'^{NG}$ we
obtain, 
\be
\begin{split}
\bar\G_n'^{NG}(\k,-\k-\q_{tot},\q_1,...\q_{n-2})
\simeq \frac{(-1)^{n-2}}{(n-3)\big(P_L(k)\big)^2}
\sum_{i=1}^{n-2}\sum_{j\neq i}^{n-2}
\frac{1}{P_L(q_j)}
\prod_{l\neq j}^{n-2}\D_{\q_l}
B_L(\k,\q_j)\;.
\end{split}
\ee 
The sum over $i$ cancels the factor $(n-3)$ from the denominator which
yields Eq.~(\ref{eq:NGosc2}). $\blacksquare$

\subsection{Damping factors}

We first compute the action of the operator ${\cal S}$ given by
Eq.~(\ref{eq:defS}) on the wiggly power spectrum
(\ref{Pwiggly}). Using 
\be
\q\cdot\nabla_\k e^{i\g\ln k/k_*}=i\g\frac{(\q\cdot\k)}{k^2}e^{i\g\ln
  k/k_*} 
\ee
we have in the leading order in $\g$,
\be 
\begin{split}
{\cal S}P^{\rm w}_L(k)
&=\int_{q\leq k_S}[dq]P_L^{\rm nw}(q)
\frac{(\k\cdot \q)^2}{q^4}\big(1-\cosh(\q\cdot \nabla_{\k'})\big)P^{\rm
  w}_L(k')\Big|_{\k'=\k}\\
&=\int_{q\leq k_S}[dq]P_L^{\rm nw}(q)
\frac{(\k\cdot \q)^2}{q^4}\bigg[1-\cos\bigg(\g\frac{\q\cdot
  \k}{k^2}\bigg)
\bigg]P^{\rm
  w}_L(k)\;.
\end{split}
\ee
Evaluating the integral over directions of the momentum $\q$ we arrive
at Eq.~(\ref{PSdamping}).

Next we consider the action of ${\cal S}$ on the oscillating bispectrum
of the form (\ref{AMBS}). Explicitly one has,
\be
\label{SBISexpl}
\begin{split}
&{\cal S}B_L(\k_1,\k_2)=\int_{q\leq k_S}[dq] P_L(q)\bigg[
\frac{(\q\cdot\k_1)^2}{q^4}\big(1-\ch(\q\cdot\nabla_{\k_1'})\big)
+\frac{(\q\cdot\k_2)^2}{q^4}\big(1-\ch(\q\cdot\nabla_{\k_2'})\big)\\
&+\frac{(\q\cdot\k_1)(\q\cdot\k_2)}{q^4}\big(1-\ch(\q\cdot\nabla_{\k_1'})
-\ch(\q\cdot\nabla_{\k_2'})+\ch(\q\cdot(\nabla_{\k_1'}-\nabla_{\k_2'}))\big)\bigg]
B_L(\k_1',\k_2')
\Big|_{\begin{smallmatrix}\k_1'=\k_1\\\k_2'=\k_2\end{smallmatrix}}
\end{split}
\ee
One makes use of 
\be
\q\cdot\nabla_{\k_1}e^{i\g\ln k_t/k_*}=\frac{i\g}{k_t}
\big((\q\cdot\hat\k_1)+(\q\cdot\hat\k_{12})\big)e^{i\g\ln k_t/k_*}
\ee
and similar relation with $\k_1$ replaced by $\k_2$. Here an overhat denotes
a unit vector along the corresponding direction. Substituting this
into (\ref{SBISexpl}) and replacing $\k_{12}$ by $-\k_3$ one casts the
result into a symmetric form,
\be
\label{SBIS1}
\begin{split} 
&{\cal S} B_L(k_1,k_2,k_3)=\\
&-\int_{q\leq k_S} [dq]P_L(q)\Bigg\{ 
 \sum_{i<j}^3\frac{(\q\cdot \k_i)(\q\cdot \k_j)}{q^4}
\left[1-\cos\left( \frac{\gamma}{k_t}
 \big( (\q\cdot(\hat\k_i-\hat\k_j)\big)\right)\right]
 \Bigg\} B_L(k_1,k_2,k_3)\;. 
 \end{split}
\ee

It remains to perform the integration over angles. To this end 
we consider the following
general integral,
\be
\label{Iintegral}
{\cal I}_{\alpha\beta}({\bf u})=\int d\hat\q \;\hat q_\alpha \hat q_\beta
\big(1-\cos(\q\cdot{\bf u})\big)\;,
\ee 
where $\hat q_{\a,\b}$ are the components of $\hat \q$ in the
Cartesian frame and ${\bf u}$ is an arbitrary vector. Due to the
rotation invariance, this must have the form,
\be
{\cal I}_{\a\b}=A(u)\hat{u}_\a\hat{u}_\b
+B(u)\delta_{\a\b}\;.
\ee 
By taking the trace of ${\cal I}_{\a\b}$ and its contraction with
$\hat u_\a\hat u_\b$ we obtain two equations on the coefficients,
\bseq
\begin{align}
&A+3B=\int d\hat\q\;\big(1-\cos(\q\cdot{\bf
  u})\big)=4\pi\big(1-j_0(qu)\big)\;,\\
&A+B=\int d\hat\q\;(\hat\q\cdot\hat{\bf u})\big(1-\cos(\q\cdot{\bf
  u})\big)=\frac{4\pi}{3}\big(1-j_0(qu)-j_2(qu)\big)\;.
\end{align}
\eseq
Thus, for the integral (\ref{Iintegral}) we find
\be
\label{Iintegral1}
{\cal I}_{\a\b}=4\pi j_2(qu)\hat{u}_\a\hat{u}_\b
+\frac{4\pi}{3}\big(1-j_0(qu)-j_2(qu)\big)\delta_{\a\b}\;.
\ee
Substitution of this result into (\ref{SBIS1}) with 
${\bf u}=\frac{\g}{k_t}(\hat\k_i-\hat\k_j)$ yields Eq.~(\ref{BSdamping}).

\section{BAO in the non-Gaussian contribution to
  the bispectrum} 
\label{app:bao}

Even if the primordial bispectrum is smooth, the linear one will
acquire an oscillatory component due to the BAO wiggles in the
transfer function, see Eq.~(\ref{earlybis}). In this Appendix we study
these wiggles using the frameworks of TSPT and discuss their damping
by large bulk flows.

Following \cite{Blas:2016sfa}, where such study was performed for
Gaussian initial conditions, we separate the linear power spectrum
into the non-wiggly and wiggly parts,
\be
P_L(k)= P_L^{\rm nw}(k) + P_L^{\rm w}(k)\;,
\ee
where the latter encapsulates the BAO.
It is characterized by oscillations in $k$ with the period
$2\pi/r_{\rm BAO}$, where $r_{\rm BAO}\simeq 110 h^{-1}\text{Mpc}$ is
the BAO scale. As the BAO component is small, we can expand all
quantities to linear order in $P_L^{\rm w}$. Thus, the linear
bispectrum becomes,
\be
B_L=B_L^{\rm nw}+B_L^{\rm w}\;,
\ee
where $B_L^{\rm nw}$ is given by Eq.~(\ref{earlybis}) with $P_L$
replaced by $P_L^{\rm nw}$ and
\be
\label{BBAO}
B_L^{\rm w}(k_1,k_2,k_3)=\bigg(
\sum_{i=1}^3\frac{P_L^{\rm w}(k_i)}{2P_L^{\rm nw}(k_i)}\bigg)
B_L^{\rm nw}(k_1,k_2,k_3)\;.
\ee
Correspondingly, the TSPT vertices, both Gaussian and non-Gaussian,
split into smooth and wiggly parts. In particular, for the cubic
non-Gaussian vertex we have,
\be
\bar\G_3'^{NG}=\bar\G_3'^{NG,\,{\rm nw}}+\bar\G_3'^{NG,\,{\rm w}}\;,
\ee
with $\bar\G_3'^{NG,\,{\rm nw}}$ constructed in the usual way
(\ref{G3NG}) out of smooth elements, and
\be
\label{G3BAO}
\bar\G_3'^{NG,\,{\rm w}}(\k_1,\k_2,\k_3)=
\frac{B_L^{\rm w}(k_1,k_2,k_3)}{P_L^{\rm nw}(k_1)P_L^{\rm nw}(k_2)
P_L^{\rm nw}(k_3)}\;.
\ee
Note that the sign in the last formula is different from (\ref{G3NG})
because of contributions coming from the expansion of the power
spectra in the denominator. Other wiggly non-Gaussian vertices are
generated from $\bar\G_3'^{NG,\,{\rm w}}$ by the recursion relation
(\ref{GNNG}). 

We can now readily use the results of Sec.~\ref{sec:IRres} to work out
the effect of IR resummation. Thus, the IR enhanced contributions to
wiggly bispectrum with three hard momenta are obtained by dressing the
wiggly elements in the two tree-level diagrams 
\be
\begin{fmffile}{bispectrumBAO_1}
\parbox{100pt}{
\begin{fmfgraph*}(80,60)
\fmfpen{thick}
\fmfleft{l1,l2}
\fmfright{r1}
\fmf{plain,label=$ $,label.side=right}{l1,b1}
\fmf{plain,label=$ $,label.side=left}{l2,b1}
\fmf{wiggly,label=$ $}{b1,r1}     
\fmfv{decor.shape=pentagram,decor.filled=empty,decor.size=10thick,label=$ $,l.a=120,l.d=.07w}{b1}
\end{fmfgraph*}}
\end{fmffile}
\quad+\quad
\begin{fmffile}{bispectrumBAO_2}
\parbox{100pt}{
\begin{fmfgraph*}(80,60)
\fmfpen{thick}
\fmfleft{l1,l2}
\fmfright{r1}
\fmf{plain,label=$ $,label.side=right}{l1,b1}
\fmf{plain,label=$ $,label.side=left}{l2,b1}
\fmf{plain,label=$ $}{b1,r1}     
\fmfv{decor.shape=pentagram,decor.filled=shaded,decor.size=10thick,label=$ $,l.a=120,l.d=.07w}{b1}
\end{fmfgraph*}}
\end{fmffile}
  \label{eq:bispectrumBAO}
\ee 
with daisy loops. Here the empty and shaded star stand for the
non-wiggly and wiggly non-Gaussian vertices, respectively; recall also
that the wiggly line denotes $P_L^{\rm w}$, whereas the plain line
stands for $P_L^{\rm nw}$. We already know the result of this dressing
from Eqs.~(\ref{PSresummed}), (\ref{BLOoper}) that provide us with its
operator form,
\be
\label{BBAOLO}
B_{\delta\delta\delta}^{NG,\,{\rm w, LO}}(\k_1,\k_2)
=g^3\bigg(\sum_{i=1}^3\frac{e^{-g^2{\cal S}}P_L^{\rm w}(k_i)}{P_L^{\rm
    nw}(k_i)}
\bigg)
B_L^{\rm nw}(\k_1,\k_2)-
g^3 e^{-g^2{\cal S}}B_L^{\rm w}(\k_1,\k_2)\;.
\ee
It is straightforward to find the action of the operator ${\cal S}$ on
$P_L^{\rm w}$ and $B_L^{\rm w}$. The result amounts to an exponential
damping of the wiggly power spectra that enter in (\ref{BBAOLO}) both
explicitly and implicitly through the expression (\ref{BBAO}),
\be
\label{replaceBAO}
P_L^{\rm w}(k)\mapsto e^{-g^2 \sigma^2 k^2} P_L^{\rm w}(k)\;.
\ee
Here
\be 
\sigma^2= \int_{q\leq k_S} \frac{dq}{6\pi^2}\,
P_L^{\rm nw}(q)\big[1-j_0\left(q r_{BAO}\right)+2j_2\left(q
    r_{BAO}\right)\big]\, 
\ee
is the same factor that appears in the context of BAO damping in the
Gaussian perturbation theory, see
e.g. \cite{Vlah:2015zda,Baldauf:2015xfa,Blas:2016sfa}. The choice of
the separation scale $k_S$ delimiting the region of soft modes
parameterizes the freedom in the resummation scheme. It is known that
good results are obtained for $k_S$ in the range 
$(0.05\div 0.2)h/\text{Mpc}$ (see a detailed discussion in
Ref.~\cite{Blas:2016sfa}). Substitution of (\ref{BBAO}),
(\ref{replaceBAO}) into (\ref{BBAOLO}) yields,
\be
\label{BBAOLO1}
B_{\delta\delta\delta}^{NG,\,{\rm w, LO}}(\eta;k_1,k_2,k_3)
=g^3(\eta)\bigg(\sum_{i=1}^3\frac{e^{-g^2(\eta)\sigma^2 k_i^2}P_L^{\rm
    w}(k_i)}{2P_L^{\rm nw}(k_i)}
\bigg)
B_L^{\rm nw}(k_1,k_2,k_3)\;.
\ee
By adding the non-wiggly part we can cast the result in a compact
form,
\be
\label{BBAOLO2}
B_{\delta\delta\delta}^{NG,\,{\rm LO}}(\eta;k_1,k_2,k_3)
=g^3(\eta)
\sqrt{\frac{P_L^{\rm res}(\eta;k_1)P_L^{\rm res}(\eta;k_2)
P_L^{\rm res}(\eta;k_3)}{P_\zeta(k_1)P_\zeta(k_2)P_\zeta(k_3)}}
B_\zeta(k_1,k_2,k_3)
\;,
\ee
where the {\it resummed linear power spectrum} is
\be
\label{eq:IRps}
P_L^{\rm res}(\e;k)=P_L^{\rm nw}(k)+e^{-g^2(\eta)\sigma^2
  k^2}P_L^{\rm w}(k)\,.
\ee
This expression has a clear intuitive interpretation. It tells us that
to obtain the leading-order IR resummed bispectrum all we have to do
is to take the usual expression,
Eq.~(\ref{earlybis}), and replace all linear matter power spectra in
it with their resummed counterparts. 

\begin{figure}[t!]
\begin{center}
  \includegraphics[width=0.5\textwidth]{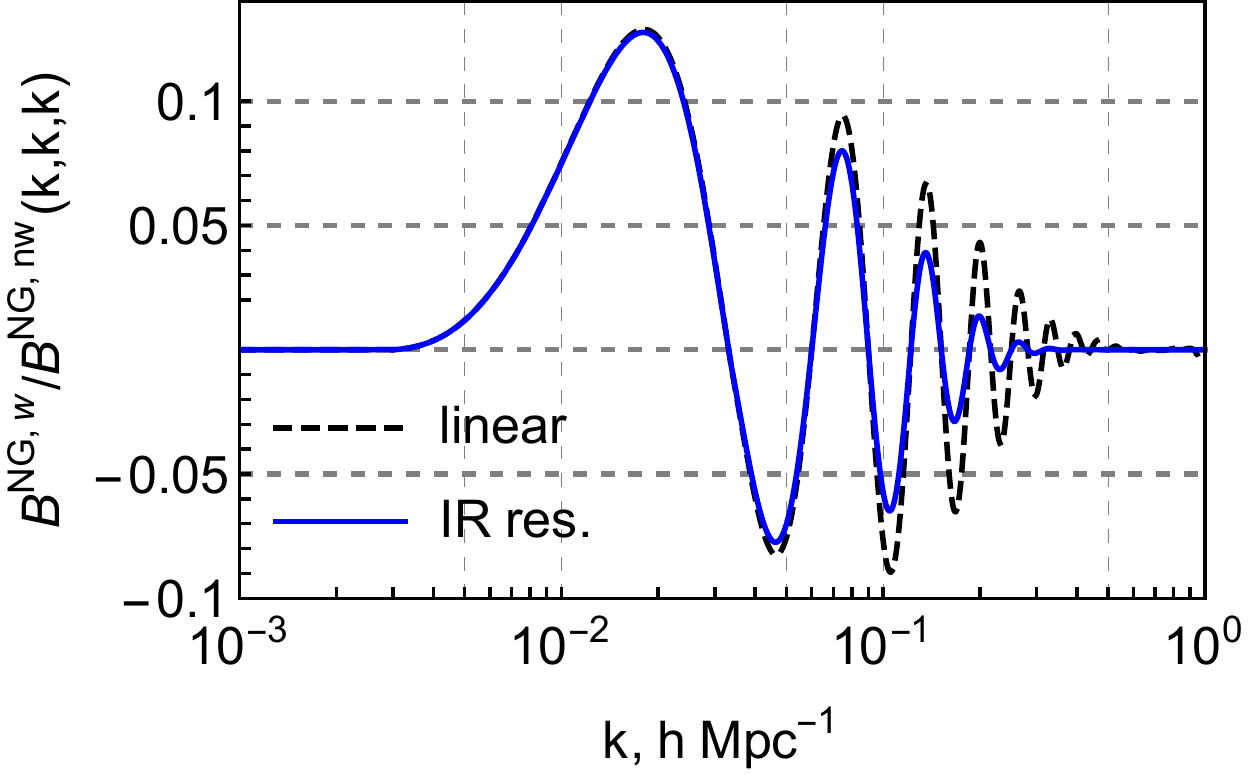}
\end{center}
\caption{\label{fig:xi}
Oscillating contribution in the non-Gaussian bispectrum due to BAO for
equilateral
configurations at redshift $z=0$. 
Blue solid (black dashed) lines show the results with
(without) IR resummation. 
The separation scale $k_S=0.2\,h$/Mpc.
}
\end{figure}

One can show that the final formula (\ref{BBAOLO2}) is also valid in
the squeezed limit. An extra contribution coming from the daisy
dressing of the non-Gaussian wiggly vertex $\bar\G_n'^{NG,\,{\rm w}}$
and corresponding to the second term in Eq.~(\ref{BLOsqueezed}) in
fact cancels against the contribution of the fish diagram composed of
the wiggly Gaussian and non-wiggly non-Gaussian vertices,
\be
\begin{fmffile}{BAOfish}
\begin{fmfgraph*}(120,50)
\fmfpen{thick}
\fmfkeep{1loop_2}
\fmfleft{l1,l2}
\fmfright{r1}
\fmfv{d.sh=circle,d.filled=shaded,d.si=7thick,label=$ $,l.a=135,l.d=.05w}{v1}
\fmf{plain,label=$\q_0$,l.side=right}{l1,v1}
\fmf{plain,label=$\k$,l.side=left}{l2,v1}
\fmf{plain,tension=1.2,label=$-\k-\q_0$}{v2,r1}
\fmf{plain,left=0.9,tension=0.5,label=$ $}{v1,v2}
\fmf{plain,left=0.9,tension=0.5,label=$ $,l.side=left}{v2,v1}
\fmfv{decor.shape=pentagram,decor.filled=empty,decor.size=7thick,label=$ $,l.a=120,l.d=.07w}{v2}
\fmfposition
\end{fmfgraph*}
\end{fmffile}
\ee
We leave the proof of this cancellation to the reader.

The effect of IR resummation on the BAO feature in the non-Gaussian
bispectrum is illustrated in Fig.~\ref{fig:xi}. We see a clear damping
of the oscillations due to large bulk flows. Of course, even apart
from the damping, the contribution of BAO in the linear bispectrum is
at most several per cent. Given the current constraints on
non-Gaussianity, measuring this feature appears unfeasible even with
futuristic surveys.

\end{document}